\documentclass[sigconf,natbib=true]{acmart}
 \usepackage{amsmath,amssymb,amsfonts,bm}
\usepackage{graphicx}
\usepackage{textcomp}
\usepackage{xcolor}
\usepackage{float}
\usepackage{array}
\usepackage{tabularx}
\usepackage{array}
\usepackage{balance}
\usepackage{marvosym}
\usepackage{multirow}
\usepackage{array}
\usepackage{enumitem}
\usepackage{booktabs}
\usepackage{algpseudocode}
\usepackage[ruled]{algorithm2e} 
\usepackage{xcolor}
\usepackage{pifont}
\usepackage{url}
\usepackage{algpseudocode}
\usepackage[ruled]{algorithm2e}

\AtBeginDocument{%
  \providecommand\BibTeX{{%
    \normalfont B\kern-0.5em{\scshape i\kern-0.25em b}\kern-0.8em\TeX}}}

\copyrightyear{2021}
\acmYear{2021}
\setcopyright{acmlicensed}\acmConference[CIKM '21]{Proceedings of the 30th ACM International Conference on Information and Knowledge Management}{November 1--5, 2021}{Virtual Event, QLD, Australia}
\acmBooktitle{Proceedings of the 30th ACM International Conference on Information and Knowledge Management (CIKM '21), November 1--5, 2021, Virtual Event, QLD, Australia}
\acmPrice{15.00}
\acmDOI{10.1145/3459637.3482388}
\acmISBN{978-1-4503-8446-9/21/11}

\begin{document}

\title{Self-Supervised Graph Co-Training for Session-based Recommendation}

\author{Xin Xia}
\affiliation{%
  \institution{The University of Queensland}
  \city{}
  \state{}
  \country{}
}
\email{x.xia@uq.edu.au}

\author{Hongzhi Yin}
\authornote{Corresponding author and having equal contribution with the first author.}

\affiliation{%
  \institution{The University of Queensland}
  \city{}
  \state{}
  \country{}
}
\email{h.yin1@uq.edu.au}

\author{Junliang Yu}
\affiliation{%
  \institution{The University of Queensland}
  \city{}
  \state{}
  \country{}
}
\email{jl.yu@uq.edu.au}
\author{Yingxia Shao}
\affiliation{%
  \institution{BUPT}
  \city{}
  \country{}
}
\email{shaoyx@bupt.edu.cn}

\author{Lizhen Cui}
\affiliation{%
  \institution{Shandong University}
  \city{}
  \state{}
  \country{}
}
\email{clz@sdu.edu.cn}

\begin{abstract}
Session-based recommendation targets next-item prediction by exploiting user behaviors within a short time period. Compared with other recommendation paradigms, session-based recommendation suffers more from the problem of data sparsity due to the very limited short-term interactions. Self-supervised learning, which can discover ground-truth samples from the raw data, holds vast potentials to tackle this problem. However, existing self-supervised recommendation models mainly rely on item/segment dropout to augment data, which are not fit for session-based recommendation because the dropout leads to sparser data, creating unserviceable self-supervision signals. In this paper, for informative session-based data augmentation, we combine self-supervised learning with co-training, and then develop a framework to enhance session-based recommendation. Technically, we first exploit the session-based graph to augment two views that exhibit the internal and external connectivities of sessions, and then we build two distinct graph encoders over the two views, which recursively leverage the different connectivity information to generate ground-truth samples to supervise each other by contrastive learning. In contrast to the dropout strategy, the proposed self-supervised graph co-training preserves the complete session information and fulfills genuine data augmentation. Extensive experiments on multiple benchmark datasets show that, session-based recommendation can be remarkably enhanced under the regime of self-supervised graph co-training, achieving the state-of-the-art performance.
\end{abstract}


\maketitle

\section{Introduction}
Recommender systems (RS) now have been pervasive and become an indispensable tool to facilitate online shopping and information delivery. Most traditional recommendation approaches share a common assumption that user behaviors are constantly recorded and available for access \cite{yu2018adaptive, yu2020enhance}. However, in some situations, recording long-term user profiles may be infeasible. For example, guests who do not log in, or users who keep personal information private do not have an accessible user profile. Session-based recommendation emerges to tackle this challenge \cite{wang2019survey}, aiming at predicting the next item only with short-term user interaction data generated in a session. Owing to its promising prospect, in the past few years, session-based recommendation has received considerable attention, and a number of models have been successively developed \cite{rendle2010factorizing,shani2005mdp,liu2018stamp,li2017neural}.\par
Early effort in this field brought Markov Chain into session-based scenarios to capture the temporal information \cite{rendle2010factorizing,shani2005mdp}. Afterwards, deep learning exhibited overwhelming advantage of modeling sequential data\cite{zhang2018discrete}, and recurrent neural networks (RNNs) became the dominant paradigm in this line of research \cite{hidasi2015session,hidasi2018recurrent}. Recently, graph neural networks (GNNs) \cite{wu2020comprehensive} have sparked heated discussions across multiple fields for its unprecedented effectiveness in solving graph-based tasks. As session-based data can also be modeled as sequence-like graphs, there also have been a proliferation of GNNs-based session-based recommendation models \cite{wu2019session,qiu2019rethinking,xia2020self,xu2019graph,pan2020star}, which outperform RNNs-based models and show decent improvements. Despite the achievements, however, these approaches are still compromised by the same issue - data sparsity. Due to the inaccessibility of the long-term user behavior data, session-based recommenders can only leverage very limited user-item interactions generated within a short session to refine the corresponding user/session representations. In most cases, these data is too few to induce an accurate user preference, leading to sub-optimal recommendation performance. \par

Self-supervised learning (SSL) \cite{liu2020self}, as an emerging learning paradigm which can discover ground-truth samples from the raw data, is considered to be an antidote to the data sparsity issue. Inspired by its great success in the areas of graph and visual representation learning \cite{jin2020self,he2020momentum}, recent advances seek to harness SSL for improving recommendation \cite{yu2021self,xia2020self,zhou2020s,xie2020contrastive}. The typical idea of applying SSL to recommendation is conducting stochastic data augmentations by randomly dropping some items/segments from the raw user-item interaction graph/sequence to create supervisory signals, which is analogous to the strategy used in masked language models like BERT \cite{devlin2018bert}. Following this line of thought, Bert4Rec \cite{sun2019bert4rec} drives a cloze objective for sequential recommendation by predicting the random masked items in the sequence with their left and right contexts. $S^{3}$-Rec \cite{zhou2020s} designs four types of pretexts to derive supervision signals from the segments, items and attributes of sequential data and then utilizes mutual information maximization to refine item representations. Similarly, CL4SRec \cite{xie2020contrastive} adopts item cropping, masking and reordering to construct different data augmentations based on sequences for contrastive learning. With such random dropout strategies, SSL is compatible with sequential recommendation. However, when it comes to session-based recommendation, the same idea may not be practicable. It should be noted that, the user interaction data generated in a session is much less than a long-term user profile in sequential recommenders. Accordingly, conducting dropout on session-based data would create sparser sequences, which could be unserviceable for improving recommendation performance. To address this problem, in this paper, we propose a novel framework which combines SSL with semi-supervised learning to create more informative self-supervision signals to enhance session-based recommendation.\par

Co-training \cite{blum1998combining}, as a classical semi-supervised learning paradigm, exploits unlabeled data to improve classifiers. The basic idea of co-training is to train two classifiers over two different data views, and then predict pseudo-labels of unlabeled instances to supervise each other in an iterative way. In our framework, we first exploit the session-item graph to construct two views (item view and session view) that exhibit the internal and external connections of sessions. Then two asymmetric graph encoders (i.e. graph convolutional networks) are built over these two views and trained under the scheme of co-training. One of them (main encoder) is for recommendation and the other acts as the auxiliary encoder to boost the former. Specifically, given a session, we regard the items as unlabeled data. In each time, one encoder predicts its possible next items and delivers them to the other encoder, respectively. By doing so, both encoders can acquire complementary information from each other. And then a contrastive objective is optimized towards refining the encoders and item representations. Meanwhile, to prevent the mode collapse (i.e. two encoders become very similar and suggest the same item), we exploit adversarial examples to encourage divergence between the two views. As this co-training regime is built upon the graph views derived from the same data source for data augmentation, and is with a contrastive objective, we name it \textit{self-supervised graph co-training}. By iterating this process, the benefits can be two-fold: (1). with the co-training proceeding, the generated item samples become more informative (a.k.a. hard examples), which can bring more useful information to each encoder compared with the dropout strategy that is only for self-discrimination; (2). the complete data of a session is preserved and two different aspects of connectivity information are exploited, generating more practicable self-supervision signals. Finally, the main encoder is significantly improved for recommendation. \par
Overall, the contributions of this paper are summarized as follows:
\begin{itemize}[leftmargin=*]
	\item We propose a novel self-supervised framework for session-based recommendation which can generate more informative and practicable self-supervision signals.
	\item The proposed framework is model-agnostic. Ideally, the architectures of the two used encoders can be diverse, which generalizes the framework to adapt to more scenarios. 
	\item Extensive experiments show that the proposed framework has overwhelming superiority over the state-of-the-art baselines and achieves statistically significant improvements on benchmark datasets. We release the code at \url{https://github.com/xiaxin1998/COTREC}.
\end{itemize}
The rest of this paper is organized as follows. Section 2 summarizes the related work of session-based recommendation and self-supervised learning. Section 3 presents the proposed framework. The experimental results are reported in Section 4. Finally, Section 5 concludes this paper.

\section{Related Work}
\subsection{Session-based Recommendation}
Early studies on session-based recommendation focused on exploiting temporal information from session data with Markov chain \cite{shani2005mdp,rendle2010factorizing,zimdars2013using,yin2016spatio}. Zimdars \textit{et al.} \cite{zimdars2013using} investigated the order of temporal data based on Markov chain and used a probability decision-tree to model sequential patterns between items. 
Shan \textit{et al.} \cite{shani2005mdp} developed a novel recommender system based on an Markov Decision Process model with appropriate initialization and generated recommendations based upon the transition probabilities between items. With the boom of deep learning, recurrent neural networks (RNNs) \cite{hochreiter1997long} have been applied to session-based recommendation models to capture sequential order between items and achieved great success \cite{zhang2014sequential,liu2016predicting}. Hidasi \textit{et al.} \cite{hidasi2015session} was the first that applied RNNs to model the whole session and introduced several modifications to vanilla RNNs such as a ranking loss function and session-parallel mini-batch training to generate more accurate recommendations. 
As a follow-up study \cite{tan2016improved}, Tan \textit{et al.} enhanced RNNs by utilizing the technique of data augmentation and handling the temporal shifts of session data. Besides, NARM \cite{li2017neural}, a neural attentive recommendation algorithm, employs a hybrid encoder with attention mechanism to model the user's sequential behavior and capture the user's main purpose in the current session. In \cite{liu2018stamp}, a short-term attention priority model is developed to capture both local and global user interests with simple multilayer perceptrons (MLPs) networks and attention mechanism. \par
Graph Neural Networks (GNNs) \cite{wu2020comprehensive} are recently introduced to session-based recommendation and exhibit great performance \cite{wu2019session,wang2020beyond, qiu2019rethinking,pan2020star, sun2020go}. Unlike RNN-based approaches, graph structure is an essential factor in graph-based methods, aiming to learn item transitions over session graphs. For example, SR-GNN \cite{wu2019session} is the first to model session sequences in session graphs and applies a gated GNN model to aggregate information between items into session representations. MGNN-SPred \cite{wang2020beyond} builds a multi-relational item graph based on all session clicks to learn global item associations and uses a gated mechanism to adaptively predict the next item. GC-SAN \cite{xu2019graph} 
dynamically constructs session-educed graphs and employs self-attention networks on the graphs to capture item dependencies via graph information aggregation. FGNN \cite{qiu2019rethinking} rethinks the sequence order of items to exploit users' intrinsic intents using GNNs. GCE-GNN \cite{wang2020global} aggregates item information from both item-level and session-level through graph convolution and self-attention mechanism. LESSR \cite{chen2020handling} proposes an edge-order preserving aggregation scheme based on GRU and a shortcut graph attention layer to address the lossy session encoding problem and effectively capture long-range dependencies, respectively. Although these graph-based methods outperform RNN-based methods, they all suffer data sparsity problem due to the limited short-term profiles in session-based scenarios.
\subsection{Self-Supervised Learning in RS}
Recently, self-supervised learning (SSL) \cite{hjelm2018learning}, as a novel machine learning paradigm which mines free labels from unlabeled data and supervises models using the generated labels, is under the spotlight. The information or intermediate representation learned from self-supervised learning are expected to carry good semantic or structural meanings and can be beneficial to a variety of downstream tasks. SSL was initially applied in the fields of visual representation learning and language modeling \cite{bachman2019learning,devlin2018bert}, where it augments the raw data through image rotation/clipping and sentence masking. Recent advances of SSL start to focus on graphs, and have received considerable attention \cite{velickovic2019deep,hu2019strategies,wu2021ssg}. DGI \cite{velickovic2019deep} maximizes mutual information between the local patch and the global graph to refine node representations, making them as the ground-truth of each other. In InfoGraph \cite{sun2019infograph}, graph-level representations encode different aspects of data by encouraging agreement between the representations of substructures with different scales (e.g., nodes, edges, triangles). Hassani \textit{et al.} \cite{hassani2020contrastive} contrasted multiple views of graphs and nodes to learn their representations. Qiu \cite{qiu2020gcc} \textit{et al.} designed a self-supervised graph neural network pre-training framework to capture the structural representations of graphs by leveraging instance discrimination and contrastive learning. \par
Inspired by the success of SSL in other areas, there are also some studies that integrate self-supervised learning into sequential recommendation \cite{zhou2020s,ma2020disentangled,xin2020self}. Bert4Rec \cite{sun2019bert4rec} transfers the cloze objective from language modeling to  sequential recommendation by predicting the random masked items in the sequence with the surrounding contexts. $S^{3}$-Rec \cite{zhou2020s} utilizes the intrinsic data correlations among attribute, item, subsequence and sequence to generate self-supervision signals and enhance the data representations via pre-training. Xie \textit{et al.} \cite{xie2020contrastive} proposed three data augmentation strategies to construct self-supervision signals from the original user behavior sequences, extracting more meaningful user patterns and encoding effective user representation. Ma \textit{et al.} \cite{ma2020disentangled} proposed a sequence-to-sequence training strategy based on latent self-supervision and disentanglement of user intention behind behavior sequences. Besides, SSL is also applied to other recommendation paradigms such as general recommendation \cite{yao2020self} and social recommendation \cite{yu2021self,yu2021socially}. Although these self-supervised methods have achieved decent improvements in recommendation performance, they are not suitable for session-based recommendation for the reason that the random dropout strategy used in these models would lead to sparser session data and unserviceable self-supervision signals. The most relevant work to ours is $S^{2}$-DHCN \cite{xia2020self} which conducts contrastive learning between representations learned over different hypergraphs by employing self-discrimination without random dropout. But it cannot learn invariant representations against the data variance for its fixed ground-truths, leading to merely slight improvements. Besides, Yu \textit{et al.} \cite{yu2021socially} recently proposed a self-supervised tri-training framework that leverages different aspects of social information to generate complementary self-supervision signals to boost recommendation. As the first work to combine SSL with multi-view semi-supervised learning for recommendation, it gives us clues about applying co-training to session-based recommendation.

\begin{figure*}[t]
	\centering
	\includegraphics[width=\textwidth]{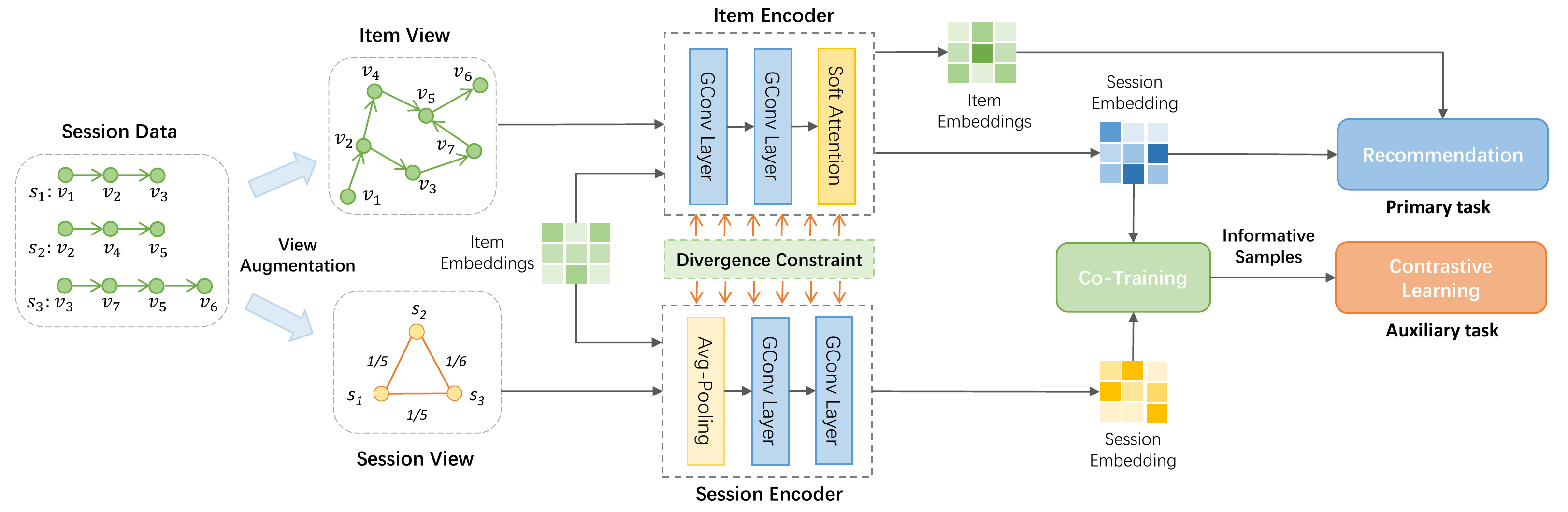}
	\caption{An overview of the proposed COTREC framework.}
	\label{figure.1}
\end{figure*}

\section{Proposed Method}
\subsection{Preliminaries}
\subsubsection{Notations.}Let $H = \{i_{1}, i_{2}, i_{3}, ... , i_{N}\}$ denote the set of items, where $N$ is the number of items. Each session is represented as a sequence $s = [i_{s,1}, i_{s,2}, i_{s,3}, ... , i_{s,m}]$ ordered by timestamps and $i_{s,k}\in H\ (1 \leq k \leq m)$ represents an interacted item of an anonymous user within the session $s$. For learning presentations, we embed each item $i\in I$ into the same space and let $\mathbf{x}_{i}^{(l)} \in \mathbb{R}^{d^{(l)}}$ denote the representation of item $i$ of dimension $d^{(l)}$ in the $l$-th layer of a deep neural network. The representation of the whole item set is denoted as $\mathbf{X}^{(l)}\in\mathbb{R}^{N \times d^{(l)}}$, and $\bm{X}^{(0)}$ is randomly initialized with uniform distribution. Each session $s$ is represented by a vector $\mathbf{s}$. The task of session-based recommendation is to predict the next item, namely $i_{s,m+1}$, for any given session $s$. Given $I$ and $s$, the output of session-based recommendation model is a ranked list $y = [y_{1}, y_{2}, y_{3}, ..., y_{N}]$ where $y_{i}\ (1 \leq i \leq N)$ is the corresponding predicted probability of item $i$. The top-\textit{K} items $(1 \leq K \leq N)$ with highest probabilities in $y$ will be selected as the recommendations. 
\subsubsection{Co-Training.}
Co-Training is a classical semi-supervised learning paradigm to exploit unlabeled data \cite{blum1998combining, da2018corec,han2020self}. Under this regime, two classifiers are separately trained on two views and then exchange confident pseudo labels of unlabeled instances to construct additional labeled training data for each other. Typically, the two views are two disjoint sets of features and can provide complementary information to each other. Blum \textit{et al.} \cite{blum1998combining} first proved that co-training can bring significant benefits when the two views are sufficient and conditionally independent. However, the required conditional dependence of two views is hard to be satisfied in many cases. To relax this assumption, Abney \textit{et al.} \cite{abney2002bootstrapping} found that weak dependence can also enable co-training success, which lifts the dependence restriction and makes co-training easily applied. Furthermore, co-training can also be applied when there is only single data representation if the data is processed by independent prediction models, such as two different classifiers \cite{xu2013survey}. It should be mentioned that there have been several attempts that combine co-training and recommendation \cite{zhang2014addressing, da2018corec}. However, these methods are all based on shallow or KNN-based models, leaving much space to be explored by the graph neural models coupled with SSL.

\subsection{Self-Supervised Graph Co-Training}
In this section, the proposed self-supervised graph \textbf{CO}-\textbf{T}raining framework for session-based \textbf{REC}ommendation (\textbf{COTREC}) is presented. The overview of COTREC is illustrated in Figure \ref{figure.1}. \par

\subsubsection{View Augmentation.}
To conduct co-training, we first derive two different views from the session data, i.e. item view and session view, by exploiting the intra- and inter-connectivity patterns of sessions. The item-view captures the item-level connectivity information while the session view encodes the session-level structural patterns. Concretely, the item view is educed by aligning all sessions. In other words, any two items ($i_{a}$ and $i_{b}$) which are connected in a session also get connected as nodes in the item view with a weighted directed edge $E_{ab}$, counting how many times they are adjacent in different sessions in the form of [$i_{a},i_{b}$]. As for the session view, two sessions ($s_{j}$ and $s_{k}$) are connected as nodes with a weighted undirected edge $E_{jk}$ obtained by using the number of shared items to divide the number of total items in the two sessions (shown in the left part of Figure \ref{figure.1}). These two views are able to provide complementary information for each other while keeping independent and exhibiting divergence to some degree, which are subject to the weak dependence constraint in \cite{abney2002bootstrapping}. To make an analogy, if we intuitively consider the session data presented in the left side of Fig.1 as the complete information, which is analogous to the whole picture in the task of image recognition, then constructing these two views corresponds to the patch clipping in visual self-supervised learning \cite{chen2020simple}. The augmented parts differ but inherit essential information from the original data, which can help learn more generalizable representations through a self-supervised task. \par

\subsubsection{Learning Graph Encoders over Augmented Views.}
After the view construction, we have two types of graphs. Although we aim to devise a model-agnostic framework that can drive a multitude of session-based neural graph recommendation models, for a concrete architecture than can fulfill the capability of the proposed self-supervised graph co-training, we construct two different graph encoders with graph convolutions over the views as the base. However, the technical details can be modified to adapt to more scenarios.

\noindent\textbf{Item View Encoding.} The item encoder with a simplified graph convolution layer for the item view is defined as:
\begin{equation}
\mathbf{X}_{I}^{(l+1)}=\hat{\mathbf{D}}_{I}^{-1} \hat{\mathbf{A}}_{I}  \mathbf{X}_{I}^{(l)}\mathbf{W}_{I}^{l},
\end{equation}
where $\hat{\mathbf{A}}_{I}=\mathbf{A}_{I}+\mathbf{I}$ and $\mathbf{I}$ is the identity matrix, $\hat{\mathbf{D}}_{I,p,p}=\sum_{q=1}^{m} \hat{\mathbf{A}}_{I,p, q}$ and $\mathbf{A}_{I}$ are the degree matrix and the adjacency matrix, $\mathbf{X}_{I}^{(l)}$ and $\mathbf{W}_{I}^{l}$ represent the $l$-th layer's item embeddings and parameter matrix of the item view, respectively. Here we do not use the non-linear function since it has been proved redundant in recommendation \cite{wu2019simplifying,yu2021self}. After passing $\mathbf{X}^{(0)}$ through $L$ graph convolution layers, we average the item embeddings obtained from each layer to be the final learned item embeddings $\mathbf{X}_{I}=\frac{1}{L+1}\sum_{l=0}^{L}\mathbf{X}^{(l)}_{I}$. Although the graph convolution can perfectly capture item connections, it cannot encode the order of items in a specific session.
Following \cite{wang2020global}, we concatenate the reversed position embeddings with the learned item representations by a learnable position matrix $\mathbf{P}_{r} = \left[\mathbf{p_{1}},\mathbf{p_{2}},\mathbf{p_{3}}, ...,\mathbf{p_{m}}\right]$, where $m$ is the length of the current session and $\mathbf{p_{m}}\in\mathbb{R}^{d}$ represents the vector of position $m$. The embedding of $t$-th item in session $s = [i_{s,1}, i_{s,2}, i_{s,3}, ..., i_{s,m}]$ is:
\begin{equation}
\mathbf{x}^{t*}_{I}=\tanh \left(\mathbf{W}_{1}\left[\mathbf{x}_{I}^{t} \| \mathbf{p}_{m-t+1}\right]+\mathbf{b}\right),
\end{equation}
where $\mathbf{W}_{1}\in\mathbb{R}^{d \times 2d}$, and $b\in\mathbb{R}^{d}$ are learnable parameters.\par

Session embeddings can be obtained by aggregating representations of items contained in that session. A soft-attention mechanism is often used in session-based recommendation methods where different items should have different priorities when learning session embeddings. We follow the strategy used in GCE-GNN \cite{wang2020global} to refine the embedding of session $s = [i_{s,1}, i_{s,2}, i_{s,3}, ..., i_{s,m}]$: 
\begin{equation}
\begin{aligned}
\alpha_{t}=\mathbf{f}^{\top} \sigma\left(\mathbf{W}_{2} \mathbf{x}_{s}+\mathbf{W}_{3} \mathbf{x}^{t*}_{I}+\mathbf{c}\right), \mathbf{\theta}_{I}=\sum_{t=1}^{m} \alpha_{t} \mathbf{x}^{t*}_{I}
\end{aligned}
\end{equation}
where $\mathbf{x}_{s}$ is the embedding of session $s$ and here it is obtained by averaging the embeddings of items within the session $s$, i.e. $\mathbf{x}_{s} = \frac{1}{m}\sum_{t=1}^{m}\mathbf{x}_{I}^{m}$. Session representation $\mathbf{\theta}_{I}$ is represented by aggregating item embeddings considering their corresponding importance. $\mathbf{f}, \mathbf{c}\in \mathbb{R}^{d}$, $\mathbf{W}_{2}\in\mathbb{R}^{d \times d}$ and $\mathbf{W}_{3}\in\mathbb{R}^{d \times d}$ are attention parameters used to learn the item weight $\alpha_{t}$.

\noindent \noindent\textbf{Session View Encoding.} The session view depicts item and session relations from the other perspective. Similarly, the session encoder conducts graph convolution on the session graph. As there are no items involved in the session graph, we first initialize the session embeddings $\bm{\Theta}_{S}^{(0)}$ by averaging the corresponding embeddings of items of each session in $\mathbf{X}^{(0)}$. And the graph convolution on session graph is defined as followed:
\begin{equation}
\mathbf{\Theta}_{S}^{(l+1)}=\hat{\mathbf{D}}_{S}^{-1} \hat{\mathbf{A}}_{S} \mathbf{\Theta}_{S}^{(l)}\mathbf{W}_{S}^{(l)},
\end{equation}
where $\hat{\mathbf{A}}_{S}=\mathbf{A}_{S}+\mathbf{I}$, $\mathbf{A}_{S}$ is the adjacency matrix, and $\hat{\mathbf{D}}_{S}$ is the corresponding degree matrix, $\mathbf{\Theta}_{S}^{(l)}$ and $\mathbf{W}_{S}^{(l)}$ represent the $l$-th layer's session embeddings and the parameter matrix, respectively. Similarly, we pass initialized session embeddings into $L$ graph convolution layers to learn session-level information. The final session representations are obtained by averaging $L$ embeddings learned at different layers, which is formulated as $\mathbf{\Theta}_{S}=\frac{1}{L+1}\sum_{l=0}^{L}\mathbf{\Theta}^{(l)}_{S}$. 

\subsubsection{Mining Self-Supervision Signals with Graph Co-Training.} In this section, we show how graph co-training mines informative self-supervision signals to enhance session-based recommendation. \par
Recall that, in the last two subsections, we build two graph encoders over two different views that can provide complementary information to each other. Therefore, it is natural to refine each encoder by exploiting the information from the other view. This can be achieved by following the regime of co-training. Given a session $p$ in the session view, we predict its positive and negative next-item samples using its representation learned over the item view:
\begin{equation}
\mathbf{y}_{I}^{p} = \text{Softmax}\left(score_{I}^{p}\right),\,\ \ \  
score_{I}^{p} = \mathbf{X}_{I}\mathbf{\theta}_{I}^{p}
\end{equation}
where $\theta_{I}^{p}$ is the representation of session $p$ in the item view, and $\mathbf{y}_{I}^{p} \in \mathbb{R}^{N}$ denotes the predicted probability of each item being recommended to session $p$ in the item view. $\theta_{I}^{p}$ can be seen as a linear classifier, and $\mathbf{X}_{I}$ is seen as the unlabeled sample set.
\par
With the computed probabilities, we can select items with the top-K highest confidence as the positive samples which act as the augmented ground-truths to supervise the session encoder. Formally, the positive sample selection is as follows:

\begin{equation}
c^{p^{+}}_{S}= \mathrm{top}\text{-}K\left(\mathbf{y}_{I}^{p} \right).
\end{equation}
As for the way to select negative samples, a straightforward idea is to take the items with the lowest scores. However, such a way can only choose easy samples which contribute little. Instead, we randomly select $K$ negative samples from the items ranked in top 10\% in $\mathbf{y}_{I}^{p}$ excluding the positives to construct $c^{p^{-}}_{S}$. These items can be seen as hard negatives which can contribute enough information, and meanwhile are less likely to fall into the set of false negatives which would mislead the learning. Analogously, we use the similar way to select informative samples for the item encoder, 
\begin{equation}
\mathbf{y}_{S}^{p} = \text{Softmax}\left(score_{S}^{p}\right),\,\ \ \  
score_{S}^{p} = \mathbf{X}^{(0)}\mathbf{\theta}_{S}^{p}
\end{equation}
where the main difference is that when selecting the top-$k$ item samples, $\bm{X}^{(0)}$ is used rather than $\bm{X}_{I}$ because the session encoder does not output convolved item embeddings. 

In each training batch, the positive and negative pseudo-labels for each session in each view are iteratively reconstructed and then are delivered to the other view as the possible next item for refining session and item representations. The intuition behind this process is that, the item samples, which receive high confidence to be the next-item in one view, should also be valuable in the other view.  Iterating this process is expected to generate more informative examples (a.k.a. harder examples). In turn, the encoders evolve under the supervision of informative examples as well, which will recursively distill harder examples. 

\subsubsection{Contrastive Learning.}
With the generated pseudo-labels, the self-supervised task used to refine encoders can be conducted through a contrastive objective. In session-based scenarios, we assume that the last clicked item in a session is the most related to the next item. Therefore, we can maximize (minimize) the agreement between the representations of the last-clicked item and the predicted items samples, accompanied with the given session representation as the session context. Formally, given a session $p$ and the predicted ground-truths, we follow InfoNCE \cite{oord2018representation}, which can maximize the lower bound of mutual information between the item pairs, as our learning objective:

\begin{equation}
\begin{split}
&\mathcal{L}_{s s l}=-\log \frac{\sum_{i \in c^{p^{+}}_{I}} \psi\left(\mathbf{x}^{last}_{I},\bm{\theta}_{I}^{p}, \mathbf{x}_{I}^{i}\right)}{\sum_{i \in c^{p^{+}}_{I}} \psi\left(\mathbf{x}^{last}_{I},\bm{\theta}_{I}^{p}, \mathbf{x}_{I}^{i}\right)+\sum_{j \in c^{p^{-}}_{I}} \psi\left(\mathbf{x}^{last}_{I},\bm{\theta}_{I}^{p}, \mathbf{x}_{I}^{j}\right)}\\
&-\log \frac{\sum_{i \in c^{p^{+}}_{S}} \psi\left(\mathbf{x}^{last}_{(0)},\bm{\theta}_{S}^{p}, \mathbf{x}_{(0)}^{i}\right)}{\sum_{c \in c^{p^{+}}_{S}} \psi\left(\mathbf{x}^{last}_{(0)},\bm{\theta}_{S}^{p}, \mathbf{x}_{(0)}^{i}\right)+\sum_{j \in c^{p^{-}}_{S}} \psi\left(\mathbf{x}^{last}_{(0)},\bm{\theta}_{S}^{p}, \mathbf{x}_{(0)}^{j}\right)}
\end{split}
\end{equation}
where $\mathbf{x}^{last}$ is the embedding of the last-clicked item of the given session, $\psi\left(\bm{x}_{1},\bm{x}_{2},\bm{x}_{3}\right) =\exp \left(f\left(\bm{x}_{1}+\bm{x}_{2},\bm{x}_{3}+\bm{x}_{2}\right) / \tau\right) $ where $\tau$ is the temperature to amplify the effect of discrimination (we empirically use 0.2 in our experiments), and $f(\cdot): \mathbb{R}^{d} \times \mathbb{R}^{d} \longmapsto \mathbb{R}$ is the discriminator function that takes two vectors as the input and then scores the agreement between them. We simply implement the discriminator by applying the cosine operation. Through the contrastive learning between the positive and negative pairs, the two views can exchange information and the last-clicked item representation can learn to infer related items with a session context, and thus item and session representations are refined.\par
In the vanilla co-training, the generated pseudo-labels are re-used in subsequent training as training labels. However, that way will make our framework less efficient because adding pseudo-labels will lead to adjacency matrix reconstruction in each iteration. Also, it may misguide the training from then on when the pseudo-labels introduce false information because the added pseudo-labels would not be removed. Therefore, in our model, we decide not to add pseudo-labels into training set in view of the above considerations. Besides, compared with the dropout based SSL methods \cite{xie2020contrastive,zhou2020s} which leverage the fragmentary sequences as self-supervision signals, our idea has the advantage of preserving the complete session information and fulfilling genuine label augmentation, and hence it is more suitable for session-based scenarios.

\subsubsection{Divergence Constraint in Co-Training.}
In our framework, the two data views for co-training are derived from the same data source by exploiting structural information in different aspects. On the one hand, this augmentation does not require two sufficient and independent data sources, which is the advantage. But on the other hand, it somehow might lead to the mode collapse problem, i.e., two encoders become similar and generated the same ground-truths when given the same session after a number of learning iterations. Therefore, it is necessary to make the two encoders differ to some degree. Following \cite{qiao2018deep}, we impose the divergence constraint on the self-supervised graph co-training regime by integrating adversarial examples into the training. \par
Theoretically, the adversarial examples targeting one encoder \cite{goodfellow2014explaining} would mislead it to generate wrong predictions. However, if the two encoders are trained to be resistant to the adversarial examples generated by each other and still output the correct predictions, we can manage to achieve the goal of keeping them different. We define the divergence constraint as follows:
\begin{equation}
\begin{split}
\mathcal{L}_{\mathrm{diff}}=&KL\left(Prob_{I}(\mathbf{X}_{I}), Prob_{S}\left(\mathbf{X}_{I}+\Delta_{adv}^{I}\right)\right)\\
&+KL\left(Prob_{S}(\mathbf{X}_{I}), Prob_{I}\left(\mathbf{X}_{I}+\Delta_{adv}^{S}\right)\right),
\end{split}
\end{equation}
where $Prob_{I}(\cdot)$ and $Prob_{S}(\cdot)$ represent the probabilities of each item to be recommended to a given session $p$, which are computed by two encoders: $Prob_{I}(\mathbf{X}_{I})=\text{Softmax}(\bm{X}_{I}\bm{\theta}_{I}^{p})$, and $Prob_{S}(\mathbf{X}_{I})=\text{Softmax}(\bm{X}_{I}\bm{\theta}_{S}^{p})$, $\Delta_{adv}^{I}$ and $\Delta_{adv}^{S}$ represent the adversarial perturbations on the item embeddings with regard to $\bm{\theta}_{I}^{p}$ and $\bm{\theta}_{S}^{p}$, respectively, and $KL(\cdot)$ denotes the KL divergence. To make it clear, $Prob_{S}\left(\mathbf{X}_{I}+\Delta_{adv}^{I}\right)$ is the probability distribution produced by the session encoder when $\bm{X}_{I}$ is perturbed by $\Delta_{adv}^{I}$. If the session encoder is immune to $\Delta_{adv}^{I}$ that is destructive to the item encoder, it will output a probability distribution similar to $Prob_{I}(\mathbf{X}_{I})$ due to shared information, resulting in a smaller loss of Eq. (9), otherwise not.  
\par
To create adversarial examples, we adopt the FGSM method proposed in \cite{goodfellow2014explaining}, which adds adversarial perturbations on model parameters through fast gradient computation. In our paper, we add adversarial perturbations on item embeddings. The perturbations $\Delta$ are updated as:
\begin{equation}
\Delta_{a d v}=\epsilon \frac{\Gamma}{\|\Gamma\|} \quad \text { where } \quad \Gamma=\frac{\partial l_{a d v}(\hat{y} \mid \mathbf{x}+\Delta)}{\partial \Delta} .
\end{equation}
$l_{a d v}(\hat{y})$ is the loss of aversarial examples and $\epsilon$ is the control parameter ($\epsilon$ is 0.5 on Diginetica and 0.2 on Tmall and RetailRocket in our experiments). 

\subsubsection{Model Optimization.}
Based on the learned representations, the score of each candidate item $i \in I$ to be recommended for a session $s$ is computed by doing inner product:
\begin{equation}
\hat{\mathbf{z}}_{i}=\mathbf{\theta}_{I}^{s\top}\mathbf{x}_{i}.
\end{equation}
Since the item view can reflect the item connectivity in a finer-grained granularity, we use the encoder over the item view as the main encoder to predict the final candidate items for recommendation. After that, a softmax function is applied:
\begin{equation}
\hat{\mathbf{y}}=\operatorname{softmax}(\hat{\mathbf{z}}).
\end{equation}
We then use cross entropy loss function to be the learning objective:
\begin{equation}
\mathcal{L}_{r}=-\sum_{i=1}^{N} \mathbf{y}_{i} \log \left(\hat{\mathbf{y}}_{i}\right)+\left(1-\mathbf{y}_{i}\right) \log \left(1-\hat{\mathbf{y}}_{i}\right).
\end{equation}
$\mathbf{y}$ is the one-hot encoding vector of the ground truth. For simplicity, we leave out the $L_{2}$ regularization terms.
Finally, we unify the recommendation task with the auxiliary SSL task. The total loss $L$ is defined as:
\begin{equation}
\mathcal{L}=\mathcal{L}_{r}+\beta \mathcal{L}_{ssl} + \alpha \mathcal{L}_{\mathrm{diff}},
\end{equation}
where $\alpha, \beta$ are hyperparameters to control the scale of the self-supervised graph co-training and view difference constraint. It should be noted that, we jointly optimize the three throughout the training. Finally, the whole procedure of COTREC is summarized in Algorithm \ref{alg:Framework}. 


\begin{algorithm}[t]
 \caption{The whole procedure of COTREC}
 \LinesNumbered 
 \label{alg:Framework}
 
 \KwIn{Sessions $\mathbf{S}$, node embeddings $\bm{X}$ \;}
 \KwOut{Recommendation lists}    
 Construct item view and session view  \;
 \For {each iteration}{
  \For {each batch}{
   Learn item and session representations through Eq. (1) - (4) \;
   \For {each session $s$}{
    Predict the probabilities of items being the positive examples in different views and obtain positive and negative examples with Eq.(5) - Eq. (7)\;    
    Compute self-supervised learning loss of two views via Eq.(8)\;
   }
  Add divergence constraint by following Eq. (9) - (10)\;
  Jointly optimize the overall objective in Eq. (14)\;
  }
   }
\end{algorithm}

\begin{table}[h]
	\renewcommand\arraystretch{1.0}
	\label{Table:1}
	\begin{center}
		\begin{tabular}{ccccc}
			\hline
			Dataset & Tmall & RetailRocket & Diginetica \\ \hline
			training sessions & 351,268 & 433,643 & 719,470\\
			test sessions & 25,898 & 15,132 & 60,858 \\
			\# of items & 40,728 & 36,968 & 43,097 \\
			average lengths & 6.69 & 5.43 & 5.12 \\
			\hline
		\end{tabular}
	\end{center}
	\caption{Dataset Statistics}
\end{table}

\begin{table*}[tp]
	\label{Table:2}
	\renewcommand\arraystretch{1.1}
	\begin{center}
		{
			\begin{tabular}{*{13}{c}}
				\toprule
				\multirow{2}{*}{Method} &
				\multicolumn{4}{c}{Tmall} & \multicolumn{4}{c}{RetailRocket} & \multicolumn{4}{c}{Diginetica} \cr
				\cmidrule(lr){2-5}\cmidrule(lr){6-9}\cmidrule(lr){10-13} & P@10 & M@10 & P@20 & M@20 & P@10 & M@10 & P@20 & M@20 & P@10 & M@10 & P@20 & M@20  \\ \hline
				
				
				FPMC &13.10 & 7.12 & 16.06 & 7.32 & 25.99 & 13.38 & 32.37 & 13.82 & 15.43 & 6.20  & 26.53 & 6.95 \\
				
				GRU4REC  & 9.47 & 5.78 & 10.93 & 5.89 & 38.35 & 23.27 & 44.01 & 23.67 & 17.93 & 7.33  & 29.45 & 8.33 \\
				
				NARM   & 19.17  & 10.42 & 23.30 & 10.70 & 42.07  & 24.88 & 50.22 & 24.59 & 35.44 & 15.13 & 49.70 & 16.17 \\
				
				STAMP  & 22.63 & 13.12  &26.47 &13.36 & 42.95 & 24.61 & 50.96 & 25.17 & 33.98 & 14.26 & 45.64 & 14.32 \\
				
				SR-GNN & 23.41  &13.45  & 27.57 & 13.72 & 43.21 & 26.07 & 50.32 & 26.57 & 36.86 & 15.52 & 50.73 & 17.59 \\
				
				GCE-GNN & 28.01 & 15.08 & 33.42 & 15.42 & - & - & - &  - & 41.16 &  18.15& $\mathbf{54.22}$ & 19.04 \\
				$S^{2}$-DHCN & 26.22 & 14.60 & 31.42 & 15.05 & 46.15 &  26.85& 53.66 & 27.30 & 39.87 & 17.53 & 53.18 & 18.44\\ \hline
				COTREC & $\mathbf{30.62}$ & $\mathbf{17.65}$ & $\mathbf{36.35}$ & $\mathbf{18.04}$ & $\mathbf{48.61}$ & $\mathbf{29.46}$ &$\mathbf{56.17}$ &$\mathbf{29.97}$ &$\mathbf{41.88}$ &$\mathbf{18.16}$  &54.18 &$\mathbf{19.07}$\\ 
				\bottomrule
		\end{tabular}}
		
	\end{center}
	\caption{Performances of all comparison methods on three datasets.}
	\vspace{-10pt}
	
\end{table*}

\section{Experiments}
\subsection{Experimental Settings}
\subsubsection{Datasets.}
We evaluate our model on three real-world benchmark datasets: \textit{Tmall}\footnote{https://tianchi.aliyun.com/dataset/dataDetail?dataId=42}, \textit{RetailRocket}\footnote{https://www.kaggle.com/retailrocket/ecommerce-dataset} and \textit{Diginetica}\footnote{http://cikm2016.cs.iupui.edu/cikm-cup/}, which are often used in session-based recommendation methods. \textit{Tmall} dataset
comes from IJCAI-15 competition, which contains anonymized
user's shopping logs on Tmall online shopping platform. \textit{RetailRocket} is
a dataset on a Kaggle contest published by an e-commerce company, which contains the user's browsing activity within six months. \textit{Diginetica} dataset describes the music listening behavior of users, and \textit{Diginetica} comes from CIKM Cup 2016. For convenience of comparing, we follow the experiment environment in \cite{wu2019session, wang2020global}. Specifically, we filter out all sessions whose length is 1 and items appearing less than 5 times. Latest data (such as, the data of last week) is set to be test set and previous data is used as training set.
Then, we augment and label the training and test datasets by using a sequence splitting method, which generates multiple labeled sequences with the corresponding labels $([i_{s,1}], i_{s,2}), ([i_{s,1}, i_{s,2}], i_{s,3}), ...,
([i_{s,1}, i_{s,2}, ..., i_{s,m-1}], i_{s,m})$ for every session $s = [i_{s,1}, i_{s,2}, i_{s,3}, ..., i_{s,m}]$. Note that the label of each sequence is the last click item in it. 
The statistics of the datasets are presented in Table 1.


				
				
				
				
				
				
        
	


\subsubsection{Baseline Methods.}
We compare COTREC with the following representative methods:
\begin{itemize}[leftmargin=*]
	\item \textbf{FPMC} \cite{rendle2010factorizing} is a sequential method based on Markov Chain. In order to adapt it to session-based recommendation, we do not consider the user latent representations when computing recommendation scores.
	\item \textbf{GRU4REC} \cite{hidasi2015session} utilizes a session-parallel mini-batch training process and adopts ranking-based loss functions to model user sequences.
	\item \textbf{NARM} \cite{li2017neural}: is a RNN-based state-of-the-art model which employs attention mechanism to capture user's main purpose and combines it with the sequential behavior to generate the recommendations.
	\item \textbf{STAMP} \cite{liu2018stamp}: adopts attention layers to replace all RNN encoders in the previous work and employs the self-attention mechanism\cite{vaswani2017attention} to enhance the session-based recommendation performance.
	\item \textbf{SR-GNN} \cite{wu2019session}:
	applies a gated graph convolutional layer to obtain item embeddings and also employs a soft-attention mechanism to compute the session embeddings.
	\item \textbf{GCE-GNN} \cite{wang2020global}: constructs two types of session-educed graphs to capture local and global information in different levels.
	\item $\mathbf{S^{2}}\textbf{-DHCN}$ \cite{xia2020self}: constructs two types of hypergraphs to learn inter- and intra-session information and uses self-supervised learning to enhance  session-based recommendation.
\end{itemize}
\subsubsection{Evaluation Metrics.}
Following \cite{wu2019session,wang2020global}, we use P@K (Precision) and MRR@K (Mean Reciprocal Rank) to evaluate the recommendation results where K is 10 or 20.
\subsubsection{Hyper-parameters Settings.}
Following previous works, we set the embedding size to 100, the batch size for mini-batch to 100, and the $L_2$ regularization to  $10^{-5}$. In our model, all parameters are initialized with the Gaussian Distribution $\mathcal{N}(\bm{0},\bm{0.1})$. We use Adam with the learning rate of 0.001 to optimize our model. For the number of layers of graph convolution on the three datasets, a three-layer setting achieves the best performance. For the baseline models, we refer to their best parameter setups reported in the original papers and directly report their results if available, since we use the same datasets and evaluation settings.

\subsection{Experimental Results}
\subsubsection{Overall Performance.}
The experimental results of overall performance are reported in Table 2, where we highlight the best results of each column in boldface. From the results, we can draw some conclusions:
\begin{itemize}[leftmargin=*]
\item Recent methods that consider temporal information (such as, GRU4REC, NARM, STAMP, SR-GNN) outperform traditional methods (FPMC) that do not, demonstrating the importance of capturing sequential dependency between items in session-based recommendation. Besides, among the methods based on RNNs-like units (RNN, LSTM, GRU), NRAM and STAMP achieve better performance than GRU4REC. This is because NRAM and STAMP not only utilize recurrent neural networks to model sequential behavior, but also utilize an attention mechanism to learn importance of each item when learning session representations. GRU4REC which only uses GRU cannot handle the shift of user preference. 

\item Graph-based baseline methods all outperform RNN-based methods, showing the great capacity of graph neural networks in modeling session data. Among them, GCE-GNN obtains higher accuracy than SR-GNN. This proves that capturing different levels of information (inter- and intra-session information) helps accurately predict user intent in session-based recommendation. $S^{2}$-DHCN
also utilize both inter- and intra-session information in hypergraph modeling and achieves promising performance. However, compared to GCE-GNN, $S^{2}$-DHCN has lower results on Tmall and Diginetica, showing that self-discrimination based SSL method is not so successful in improving session-based recommendation performance, which is in line with our motivation.
\item Our proposed COTREC almost outperforms all the baselines on all the datasets. Particularly, it beats other models by a large margin on Tmall, showing the effectiveness of the self-supervised grpah co-training when applied to real e-commerce data. Compared with the other self-supervised model $S^{2}$-DHCN, the advantage is also obvious. Considering that $S^{2}$-DHCN and COTREC both have a two-branch architecture, we think that the improvements mainly derive from the multi-instance contrastive learning in Eq. (8) while $S^{2}$-DHCN only conducts self-discrimination contrastive learning. Compared with another strong baseline GCE-GNN, COTREC is competitive in terms of both performance and efficiency. Although GCE-GNN can achieve comparable results on Diginetica, its more complex structure makes it suffer from the out-of-memory problem when performing on RetailRocket on our RTX 2080 Ti GPU. Besides, its performance is much lower than that of COTREC on Tmall.     

\end{itemize}

\subsubsection{Ablation Study.}
In this section, to investigate the contribution of each component in our model, we develop four variant versions of COTREC: \textbf{COTREC-base}, \textbf{base-NP}, \textbf{base-NA}, \textbf{COTREC-ND}, and we compare the four variants with the complete COTREC model on \textit{Tmall} and \textit{Diginetica}.
In \textbf{COTREC-base}, we only use the item view to model session data, removing the session view and the self-supervised graph co-training. In \textbf{base-NP}, we remove the reversed position embeddings.
\textbf{base-NA} means that we remove the soft-attention mechanism and replace it with averaging item representations as the representation of each session. In \textbf{COTREC-ND}, we only use self-supervised co-training without the divergence constraint. We show the results of these four variants in Figure 2. \par
From Figure 2, we can observe that each component consistently contributes on both datasets. The self-supervised co-training improves the base model the most, serving as the driving force of the performance improvement. When removing the self-supervised co-training, we can observe a remarkable performance drop on both the two metrics. Besides, the divergence constraint is effective to prevent mode collapse in co-training process. Without this module, the performance of COTREC is even worse than that of the base on Diginetica. Note that on Tmall, base-NP outperforms COTREC-base, proving that a strict temporal order of items may negatively influence the performance in some cases, which is in line with the our previous observation in $S^{2}$-DHCN. According to the results of base-NA, it is shown that learning different item importance across sessions is better than directly averaging representations of contained items for learning session representations in session-based recommendation.
\begin{figure}[t]
	\centering
	\includegraphics[width=0.45\textwidth]{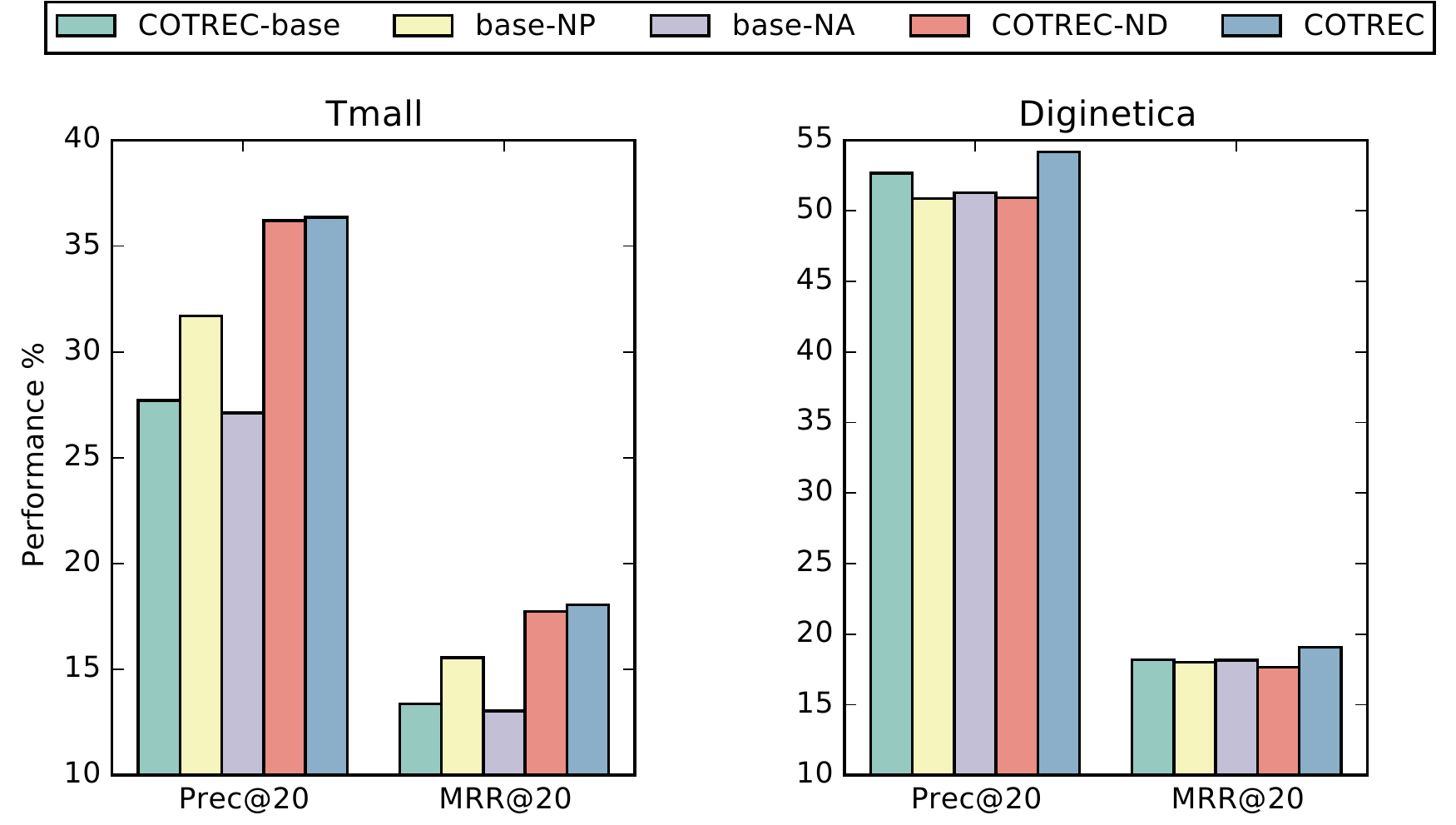}
	\caption{Ablation Study.}
	\label{figure.2}
	\vspace{-10pt}
\end{figure}

\begin{table}[h]
	\renewcommand\tabcolsep{10.0pt}
	\label{Table:3}
	\begin{center}
		\begin{tabular}{ccccc}
			\toprule
			\multirow{2}{*}{Method} &
			\multicolumn{2}{c}{Tmall} & \multicolumn{2}{c}{Diginetica} \cr
			\cmidrule(lr){2-3}\cmidrule(lr){4-5}
			 & P@20 & M@20 & P@20 & M@20 \\ \hline

			COTREC-base & 27.71 & 13.36 & 52.66&18.19 \\
			base-DHCN & 32.94 & 16.22 & 52.02 &17.70  \\
			base-MASK &32.54 & 16.37 & 51.61 & 17.64 \\
			COTREC & 36.35 & 18.04  & 54.18 & 19.07 \\
			\hline
		\end{tabular}
	\end{center}
	\caption{Comparisons of Different SSL Methods.}
	\vspace{-20pt}
\end{table}

\subsubsection{Comparison with Different SSL Methods.}
To further investigate the effectiveness of the proposed self-supervised graph co-training, we also compare it with other different SSL methods that are based on self-discrimination and random dropout to generate self-supervision signals on Tmall and Diginetica. The first is the method proposed in $S^{2}$-DHCN. DHCN proposes to capture item-level and session-level information and maximize mutual information between the session representations learned at the two levels. Positive examples are two types of session representation of the same session, whereas negative pairs are representations of different sessions. The second compared SSL method is based on the item mask, which is an often used strategy where some items are randomly dropped in each session. The generated new session can be the positive example and other sessions can be negative samples. For a fair comparison, we employ these SSL strategies on the base model of COTREC. So we name the three as \textbf{base-COTREC}, \textbf{base-DHCN}, \textbf{base-MASK}. Besides, these SSL methods are used to establish auxiliary tasks in the model optimization and we use a hyperparameter to control the magnitude of SSL. Finally, we use grid-search to adjust the parameter to ensure the best performances of them, and the best results are shown in Table 3.\par
From Table 3, we can see that, only the self-supervised graph co-training can boost the recommendation performance on both datasets and it also achieves the best performance, while the other two methods can only take effect on Tmall, demonstrating that self-supervised graph co-training is more effective than self-discrimination and dropout-based methods. We also find that the strategy of item mask is the least effective in most cases, proving that masked sub-sequences can only generate sub-optimal self-supervision signals in the scenario of session-based recommendation due to the very limited behaviors.  

\begin{figure}[t]
	\centering
	\includegraphics[width=0.5\textwidth]{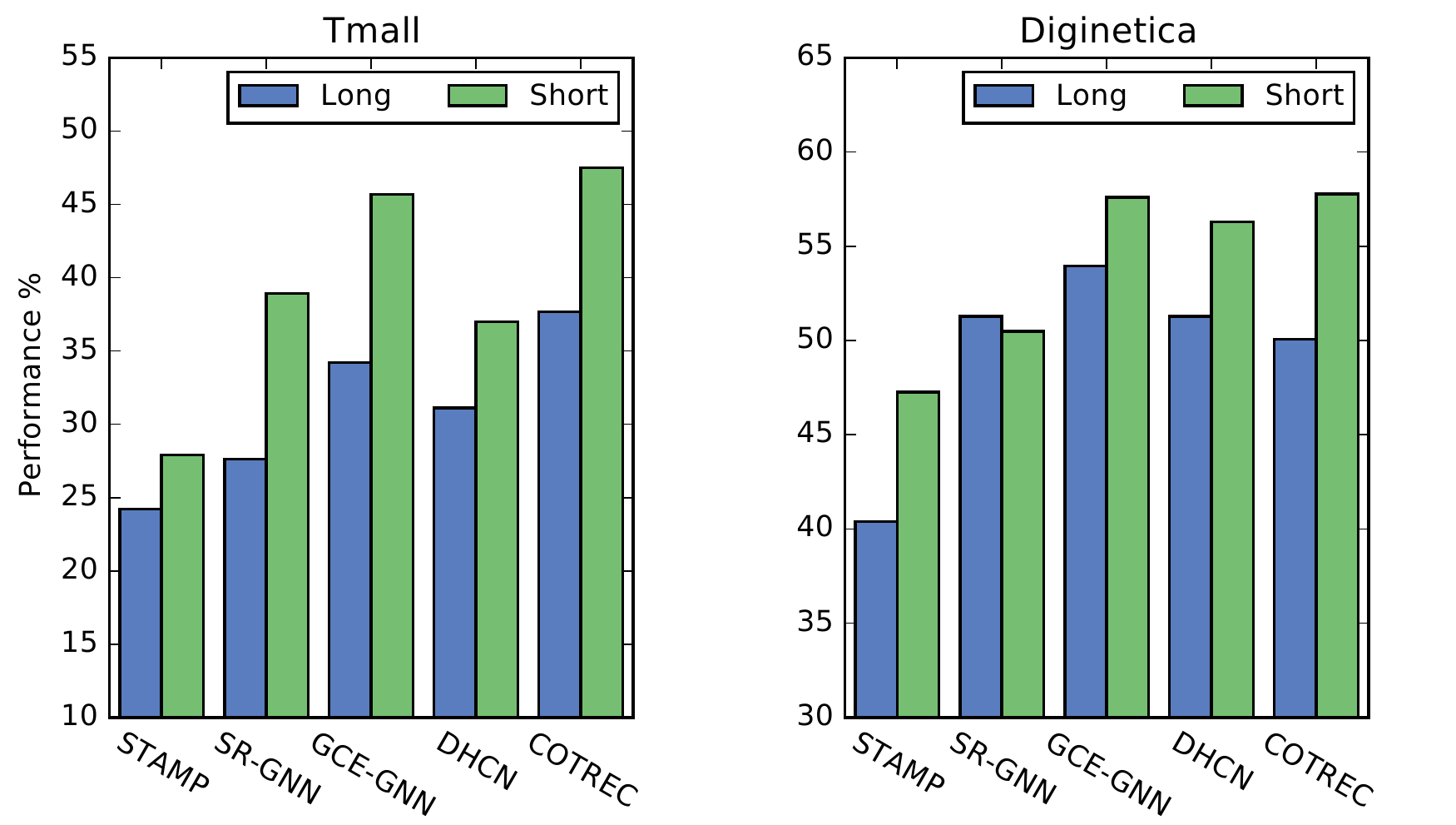}
	\caption{P@20 results on Long and Short.}
	\label{figure.3}
\end{figure}

\begin{figure}[t]
	\centering
	\includegraphics[width=0.5\textwidth]{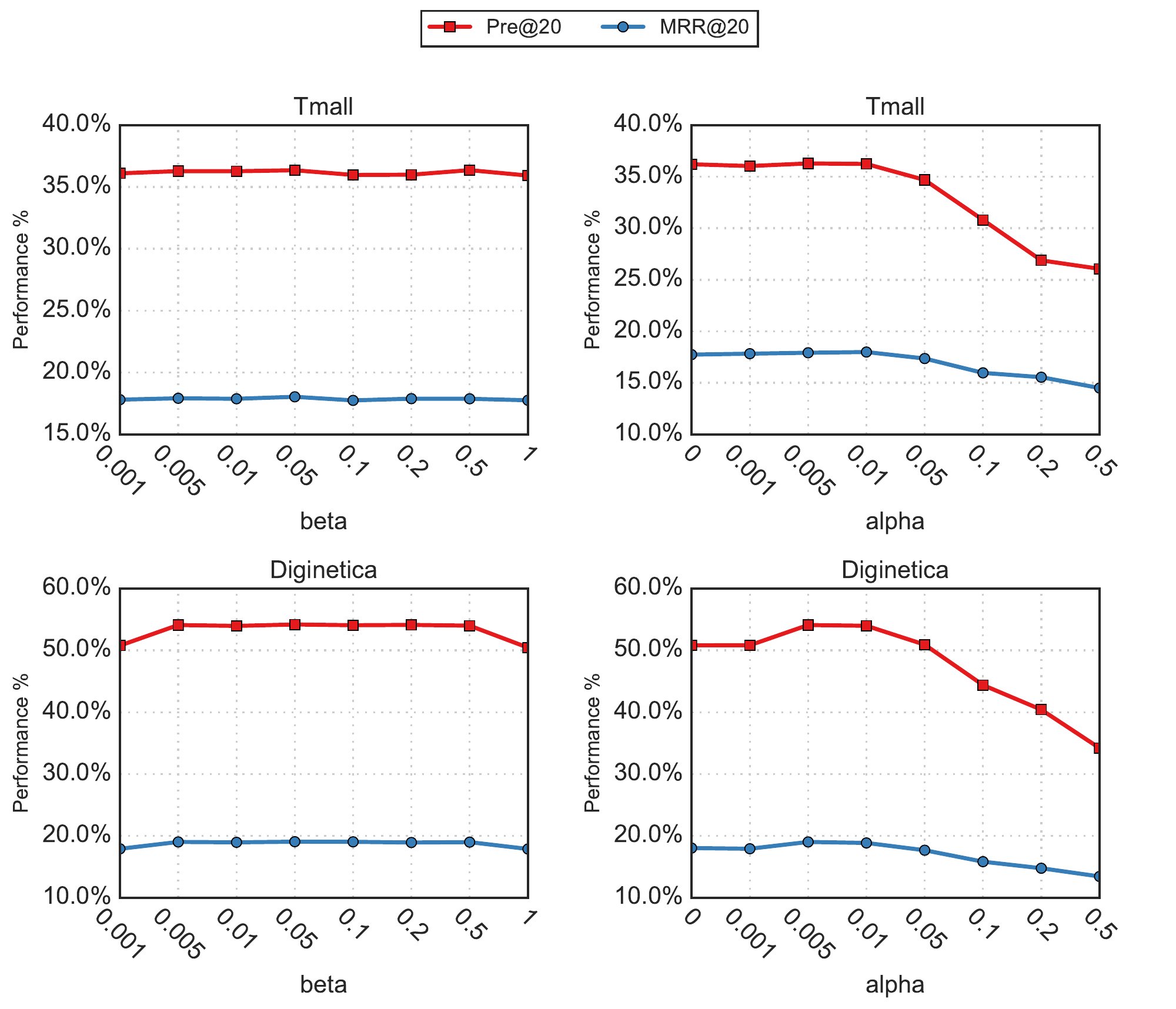}
	\caption{Hyperparameter Analysis.}
	\label{figure.4}
\end{figure}
\begin{figure}[t]
	\centering
	\includegraphics[width=0.5\textwidth]{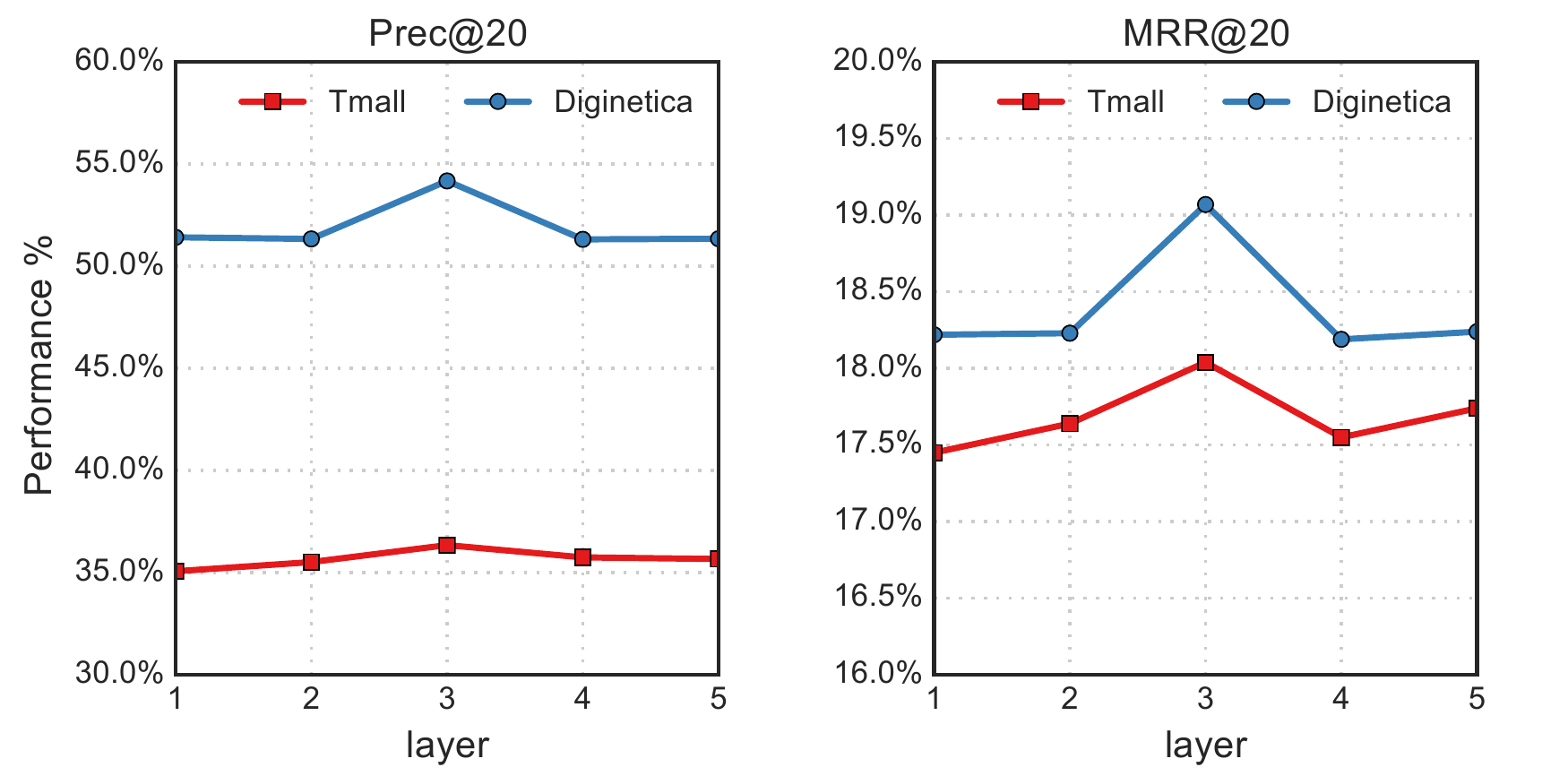}
	\caption{The impacts of the number of layer.}
	\label{figure.5}
\end{figure}

\subsubsection{Handling Different Session Lengths.} In real world situations, sessions with various lengths are common, so it is interesting to know how stable our COTREC as well as the baseline models are when dealing with them, and it is also a critical indicator for production environments. To evaluate this, we follow \cite{liu2018stamp, wu2019session} to split the sessions of \textit{Tmall} and \textit{Diginetica} into two groups with different lengths and name them as \textbf{Short} and \textbf{Long}.
Short contains sessions whose lengths are less than or equal to 5, while Long contains sessions whose lengths are larger than 5. Then, we compare the performance of COTREC with some representative baselines, i.e., STAMP, SR-GNN, GCE-GNN and $S^{2}$-DHCN in terms of Prec@20 on Short and Long. Results in Figure \ref{figure.3} show that COTREC almost outperforms all the baseline models on both datasets with different session lengths. It demonstrates the adaptability of COTREC in real-world session-based recommendation. Besides, it is shown that the performance on the short sessions is better than that on the long sessions.

\subsubsection{The Impact of Hyperparameters.}
In COTREC, we have two hyperparameters to control the magnitude of the SSL taks and the effect of the divergence constraint, i.e. $\beta$ and $\alpha$. To investigate the influence of them, we report the performance with a set of representative $\beta$ values in \{ 0.001, 0.005, 0.01, 0.05, 0.1, 0.2, 0.5, 1\} and $\alpha$ values in \{0, 0.001, 0.005, 0.01, 0.05, 0.1, 0.2, 0.5\} on Tmall and Diginetica. We fix the other parameter as 0.005 when investigating $\beta$ or $\alpha$. According to the results in Figure 4, our model achieves the best performance when jointly trained with the SSL taks and the divergence constraint. On both Tmall and Diginetica, the best setting is $\beta=0.05$ and $\alpha=0.005$. When using large $\alpha$, a huge performance drop is observed, demonstrating that when there is large divergence between two encoder, it is hard for them to supervise each other. 

\subsubsection{The impact of the number of layers.}
To investigate the impact of the number of layers in graph convolution network, we range the number of layers in \{1, 2, 3, 4, 5\}. According to the results in Figure 5, we can see that for both \textit{Tmall} and \textit{Diginetica}, a three-layer setting achieves the best performance. When the number becomes larger, performance will drop due to the over-smoothing issue. Besides, an obvious performance fluctuation is observed on \textit{Diginetica}.

\section{Conclusion}
Self-supervised learning is an emerging machine learning paradigm which exploits unlabeled data by generating ground-truth labels from the raw data itself and recently has been utilized in many fields to enhance deep learning models. Existing SSL-based recommendation methods usually adopt random dropout-based self-discrimination to generate self-supervision signals. However, we argue that it cannot adapt to session-based recommendation because it would create sparser data and cannot leverage informative self-supervision signals from other entities. In this paper, we design a self-supervised graph co-training framework to address this issue. In our framework, co-training can iteratively selects evolving pseudo-labels as informative self-supervision examples for each view to improve the session-based recommendation. Extensive experiments and empirical studies demonstrate the effectiveness of our framework and show its superiority over other recent baselines.
\section{Acknowledgment}
This work was supported by ARC Discovery Project (Grant No. DP190101985),ARC Future Fellowship (FT210100624), National Natural Science Foundation of China (No. U1936104), CCF- Baidu Open Fund, and The Fundamental Research Funds for the Central Universities 2020RC25.
\bibliographystyle{ACM-Reference-Format}
\bibliography{ref}


\begin{thebibliography}{58}


\ifx \showCODEN    \undefined \def \showCODEN     #1{\unskip}     \fi
\ifx \showDOI      \undefined \def \showDOI       #1{#1}\fi
\ifx \showISBNx    \undefined \def \showISBNx     #1{\unskip}     \fi
\ifx \showISBNxiii \undefined \def \showISBNxiii  #1{\unskip}     \fi
\ifx \showISSN     \undefined \def \showISSN      #1{\unskip}     \fi
\ifx \showLCCN     \undefined \def \showLCCN      #1{\unskip}     \fi
\ifx \shownote     \undefined \def \shownote      #1{#1}          \fi
\ifx \showarticletitle \undefined \def \showarticletitle #1{#1}   \fi
\ifx \showURL      \undefined \def \showURL       {\relax}        \fi
\providecommand\bibfield[2]{#2}
\providecommand\bibinfo[2]{#2}
\providecommand\natexlab[1]{#1}
\providecommand\showeprint[2][]{arXiv:#2}

\bibitem[\protect\citeauthoryear{Abney}{Abney}{2002}]%
        {abney2002bootstrapping}
\bibfield{author}{\bibinfo{person}{Steven Abney}.}
  \bibinfo{year}{2002}\natexlab{}.
\newblock \showarticletitle{Bootstrapping}. In
  \bibinfo{booktitle}{\emph{Proceedings of the 40th annual meeting of the
  Association for Computational Linguistics}}. \bibinfo{pages}{360--367}.
\newblock


\bibitem[\protect\citeauthoryear{Bachman, Hjelm, and Buchwalter}{Bachman
  et~al\mbox{.}}{2019}]%
        {bachman2019learning}
\bibfield{author}{\bibinfo{person}{Philip Bachman}, \bibinfo{person}{R~Devon
  Hjelm}, {and} \bibinfo{person}{William Buchwalter}.}
  \bibinfo{year}{2019}\natexlab{}.
\newblock \showarticletitle{Learning representations by maximizing mutual
  information across views}. In \bibinfo{booktitle}{\emph{Advances in Neural
  Information Processing Systems}}. \bibinfo{pages}{15535--15545}.
\newblock


\bibitem[\protect\citeauthoryear{Blum and Mitchell}{Blum and Mitchell}{1998}]%
        {blum1998combining}
\bibfield{author}{\bibinfo{person}{Avrim Blum} {and} \bibinfo{person}{Tom
  Mitchell}.} \bibinfo{year}{1998}\natexlab{}.
\newblock \showarticletitle{Combining labeled and unlabeled data with
  co-training}. In \bibinfo{booktitle}{\emph{Proceedings of the eleventh annual
  conference on Computational learning theory}}. \bibinfo{pages}{92--100}.
\newblock


\bibitem[\protect\citeauthoryear{Chen, Kornblith, Norouzi, and Hinton}{Chen
  et~al\mbox{.}}{2020}]%
        {chen2020simple}
\bibfield{author}{\bibinfo{person}{Ting Chen}, \bibinfo{person}{Simon
  Kornblith}, \bibinfo{person}{Mohammad Norouzi}, {and}
  \bibinfo{person}{Geoffrey Hinton}.} \bibinfo{year}{2020}\natexlab{}.
\newblock \showarticletitle{A simple framework for contrastive learning of
  visual representations}. In \bibinfo{booktitle}{\emph{International
  conference on machine learning}}. PMLR, \bibinfo{pages}{1597--1607}.
\newblock


\bibitem[\protect\citeauthoryear{Chen and Wong}{Chen and Wong}{2020}]%
        {chen2020handling}
\bibfield{author}{\bibinfo{person}{Tianwen Chen} {and} \bibinfo{person}{Raymond
  Chi-Wing Wong}.} \bibinfo{year}{2020}\natexlab{}.
\newblock \showarticletitle{Handling Information Loss of Graph Neural Networks
  for Session-based Recommendation}. In \bibinfo{booktitle}{\emph{Proceedings
  of the 26th ACM SIGKDD International Conference on Knowledge Discovery \&
  Data Mining}}. \bibinfo{pages}{1172--1180}.
\newblock


\bibitem[\protect\citeauthoryear{Da~Costa, Manzato, and Campello}{Da~Costa
  et~al\mbox{.}}{2018}]%
        {da2018corec}
\bibfield{author}{\bibinfo{person}{Arthur~F Da~Costa},
  \bibinfo{person}{Marcelo~G Manzato}, {and} \bibinfo{person}{Ricardo~JGB
  Campello}.} \bibinfo{year}{2018}\natexlab{}.
\newblock \showarticletitle{CoRec: a co-training approach for recommender
  systems}. In \bibinfo{booktitle}{\emph{Proceedings of the 33rd Annual ACM
  Symposium on Applied Computing}}. \bibinfo{pages}{696--703}.
\newblock


\bibitem[\protect\citeauthoryear{Devlin, Chang, Lee, and Toutanova}{Devlin
  et~al\mbox{.}}{2018}]%
        {devlin2018bert}
\bibfield{author}{\bibinfo{person}{Jacob Devlin}, \bibinfo{person}{Ming-Wei
  Chang}, \bibinfo{person}{Kenton Lee}, {and} \bibinfo{person}{Kristina
  Toutanova}.} \bibinfo{year}{2018}\natexlab{}.
\newblock \showarticletitle{Bert: Pre-training of deep bidirectional
  transformers for language understanding}.
\newblock \bibinfo{journal}{\emph{arXiv preprint arXiv:1810.04805}}
  (\bibinfo{year}{2018}).
\newblock


\bibitem[\protect\citeauthoryear{Goodfellow, Shlens, and Szegedy}{Goodfellow
  et~al\mbox{.}}{2014}]%
        {goodfellow2014explaining}
\bibfield{author}{\bibinfo{person}{Ian~J Goodfellow}, \bibinfo{person}{Jonathon
  Shlens}, {and} \bibinfo{person}{Christian Szegedy}.}
  \bibinfo{year}{2014}\natexlab{}.
\newblock \showarticletitle{Explaining and harnessing adversarial examples}.
\newblock \bibinfo{journal}{\emph{arXiv preprint arXiv:1412.6572}}
  (\bibinfo{year}{2014}).
\newblock


\bibitem[\protect\citeauthoryear{Han, Xie, and Zisserman}{Han
  et~al\mbox{.}}{2020}]%
        {han2020self}
\bibfield{author}{\bibinfo{person}{Tengda Han}, \bibinfo{person}{Weidi Xie},
  {and} \bibinfo{person}{Andrew Zisserman}.} \bibinfo{year}{2020}\natexlab{}.
\newblock \showarticletitle{Self-supervised co-training for video
  representation learning}.
\newblock \bibinfo{journal}{\emph{arXiv preprint arXiv:2010.09709}}
  (\bibinfo{year}{2020}).
\newblock


\bibitem[\protect\citeauthoryear{Hassani and Khasahmadi}{Hassani and
  Khasahmadi}{2020}]%
        {hassani2020contrastive}
\bibfield{author}{\bibinfo{person}{Kaveh Hassani} {and}
  \bibinfo{person}{Amir~Hosein Khasahmadi}.} \bibinfo{year}{2020}\natexlab{}.
\newblock \showarticletitle{Contrastive Multi-View Representation Learning on
  Graphs}.
\newblock \bibinfo{journal}{\emph{arXiv preprint arXiv:2006.05582}}
  (\bibinfo{year}{2020}).
\newblock


\bibitem[\protect\citeauthoryear{He, Fan, Wu, Xie, and Girshick}{He
  et~al\mbox{.}}{2020}]%
        {he2020momentum}
\bibfield{author}{\bibinfo{person}{Kaiming He}, \bibinfo{person}{Haoqi Fan},
  \bibinfo{person}{Yuxin Wu}, \bibinfo{person}{Saining Xie}, {and}
  \bibinfo{person}{Ross Girshick}.} \bibinfo{year}{2020}\natexlab{}.
\newblock \showarticletitle{Momentum contrast for unsupervised visual
  representation learning}. In \bibinfo{booktitle}{\emph{Proceedings of the
  IEEE/CVF Conference on Computer Vision and Pattern Recognition}}.
  \bibinfo{pages}{9729--9738}.
\newblock


\bibitem[\protect\citeauthoryear{Hidasi and Karatzoglou}{Hidasi and
  Karatzoglou}{2018}]%
        {hidasi2018recurrent}
\bibfield{author}{\bibinfo{person}{Bal{\'a}zs Hidasi} {and}
  \bibinfo{person}{Alexandros Karatzoglou}.} \bibinfo{year}{2018}\natexlab{}.
\newblock \showarticletitle{Recurrent neural networks with top-k gains for
  session-based recommendations}. In \bibinfo{booktitle}{\emph{Proceedings of
  the 27th ACM International Conference on Information and Knowledge
  Management}}. \bibinfo{pages}{843--852}.
\newblock


\bibitem[\protect\citeauthoryear{Hidasi, Karatzoglou, Baltrunas, and
  Tikk}{Hidasi et~al\mbox{.}}{2015}]%
        {hidasi2015session}
\bibfield{author}{\bibinfo{person}{Bal{\'a}zs Hidasi},
  \bibinfo{person}{Alexandros Karatzoglou}, \bibinfo{person}{Linas Baltrunas},
  {and} \bibinfo{person}{Domonkos Tikk}.} \bibinfo{year}{2015}\natexlab{}.
\newblock \showarticletitle{Session-based recommendations with recurrent neural
  networks}.
\newblock \bibinfo{journal}{\emph{arXiv preprint arXiv:1511.06939}}
  (\bibinfo{year}{2015}).
\newblock


\bibitem[\protect\citeauthoryear{Hjelm, Fedorov, Lavoie-Marchildon, Grewal,
  Bachman, Trischler, and Bengio}{Hjelm et~al\mbox{.}}{2018}]%
        {hjelm2018learning}
\bibfield{author}{\bibinfo{person}{R~Devon Hjelm}, \bibinfo{person}{Alex
  Fedorov}, \bibinfo{person}{Samuel Lavoie-Marchildon}, \bibinfo{person}{Karan
  Grewal}, \bibinfo{person}{Phil Bachman}, \bibinfo{person}{Adam Trischler},
  {and} \bibinfo{person}{Yoshua Bengio}.} \bibinfo{year}{2018}\natexlab{}.
\newblock \showarticletitle{Learning deep representations by mutual information
  estimation and maximization}.
\newblock \bibinfo{journal}{\emph{arXiv preprint arXiv:1808.06670}}
  (\bibinfo{year}{2018}).
\newblock


\bibitem[\protect\citeauthoryear{Hochreiter and Schmidhuber}{Hochreiter and
  Schmidhuber}{1997}]%
        {hochreiter1997long}
\bibfield{author}{\bibinfo{person}{Sepp Hochreiter} {and}
  \bibinfo{person}{J{\"u}rgen Schmidhuber}.} \bibinfo{year}{1997}\natexlab{}.
\newblock \showarticletitle{Long short-term memory}.
\newblock \bibinfo{journal}{\emph{Neural computation}} \bibinfo{volume}{9},
  \bibinfo{number}{8} (\bibinfo{year}{1997}), \bibinfo{pages}{1735--1780}.
\newblock


\bibitem[\protect\citeauthoryear{Hu, Liu, Gomes, Zitnik, Liang, Pande, and
  Leskovec}{Hu et~al\mbox{.}}{2019}]%
        {hu2019strategies}
\bibfield{author}{\bibinfo{person}{Weihua Hu}, \bibinfo{person}{Bowen Liu},
  \bibinfo{person}{Joseph Gomes}, \bibinfo{person}{Marinka Zitnik},
  \bibinfo{person}{Percy Liang}, \bibinfo{person}{Vijay Pande}, {and}
  \bibinfo{person}{Jure Leskovec}.} \bibinfo{year}{2019}\natexlab{}.
\newblock \showarticletitle{Strategies for pre-training graph neural networks}.
\newblock \bibinfo{journal}{\emph{arXiv preprint arXiv:1905.12265}}
  (\bibinfo{year}{2019}).
\newblock


\bibitem[\protect\citeauthoryear{Jin, Derr, Liu, Wang, Wang, Liu, and Tang}{Jin
  et~al\mbox{.}}{2020}]%
        {jin2020self}
\bibfield{author}{\bibinfo{person}{Wei Jin}, \bibinfo{person}{Tyler Derr},
  \bibinfo{person}{Haochen Liu}, \bibinfo{person}{Yiqi Wang},
  \bibinfo{person}{Suhang Wang}, \bibinfo{person}{Zitao Liu}, {and}
  \bibinfo{person}{Jiliang Tang}.} \bibinfo{year}{2020}\natexlab{}.
\newblock \showarticletitle{Self-supervised learning on graphs: Deep insights
  and new direction}.
\newblock \bibinfo{journal}{\emph{arXiv preprint arXiv:2006.10141}}
  (\bibinfo{year}{2020}).
\newblock


\bibitem[\protect\citeauthoryear{Li, Ren, Chen, Ren, Lian, and Ma}{Li
  et~al\mbox{.}}{2017}]%
        {li2017neural}
\bibfield{author}{\bibinfo{person}{Jing Li}, \bibinfo{person}{Pengjie Ren},
  \bibinfo{person}{Zhumin Chen}, \bibinfo{person}{Zhaochun Ren},
  \bibinfo{person}{Tao Lian}, {and} \bibinfo{person}{Jun Ma}.}
  \bibinfo{year}{2017}\natexlab{}.
\newblock \showarticletitle{Neural attentive session-based recommendation}. In
  \bibinfo{booktitle}{\emph{Proceedings of the 2017 ACM on Conference on
  Information and Knowledge Management}}. \bibinfo{pages}{1419--1428}.
\newblock


\bibitem[\protect\citeauthoryear{Liu, Wu, Wang, and Tan}{Liu
  et~al\mbox{.}}{2016}]%
        {liu2016predicting}
\bibfield{author}{\bibinfo{person}{Qiang Liu}, \bibinfo{person}{Shu Wu},
  \bibinfo{person}{Liang Wang}, {and} \bibinfo{person}{Tieniu Tan}.}
  \bibinfo{year}{2016}\natexlab{}.
\newblock \showarticletitle{Predicting the next location: A recurrent model
  with spatial and temporal contexts}. In \bibinfo{booktitle}{\emph{Thirtieth
  AAAI conference on artificial intelligence}}.
\newblock


\bibitem[\protect\citeauthoryear{Liu, Zeng, Mokhosi, and Zhang}{Liu
  et~al\mbox{.}}{2018}]%
        {liu2018stamp}
\bibfield{author}{\bibinfo{person}{Qiao Liu}, \bibinfo{person}{Yifu Zeng},
  \bibinfo{person}{Refuoe Mokhosi}, {and} \bibinfo{person}{Haibin Zhang}.}
  \bibinfo{year}{2018}\natexlab{}.
\newblock \showarticletitle{STAMP: short-term attention/memory priority model
  for session-based recommendation}. In \bibinfo{booktitle}{\emph{Proceedings
  of the 24th ACM SIGKDD International Conference on Knowledge Discovery \&
  Data Mining}}. \bibinfo{pages}{1831--1839}.
\newblock


\bibitem[\protect\citeauthoryear{Liu, Zhang, Hou, Wang, Mian, Zhang, and
  Tang}{Liu et~al\mbox{.}}{2020}]%
        {liu2020self}
\bibfield{author}{\bibinfo{person}{Xiao Liu}, \bibinfo{person}{Fanjin Zhang},
  \bibinfo{person}{Zhenyu Hou}, \bibinfo{person}{Zhaoyu Wang},
  \bibinfo{person}{Li Mian}, \bibinfo{person}{Jing Zhang}, {and}
  \bibinfo{person}{Jie Tang}.} \bibinfo{year}{2020}\natexlab{}.
\newblock \showarticletitle{Self-supervised learning: Generative or
  contrastive}.
\newblock \bibinfo{journal}{\emph{arXiv preprint arXiv:2006.08218}}
  \bibinfo{volume}{1}, \bibinfo{number}{2} (\bibinfo{year}{2020}).
\newblock


\bibitem[\protect\citeauthoryear{Ma, Zhou, Yang, Cui, Wang, and Zhu}{Ma
  et~al\mbox{.}}{2020}]%
        {ma2020disentangled}
\bibfield{author}{\bibinfo{person}{Jianxin Ma}, \bibinfo{person}{Chang Zhou},
  \bibinfo{person}{Hongxia Yang}, \bibinfo{person}{Peng Cui},
  \bibinfo{person}{Xin Wang}, {and} \bibinfo{person}{Wenwu Zhu}.}
  \bibinfo{year}{2020}\natexlab{}.
\newblock \showarticletitle{Disentangled Self-Supervision in Sequential
  Recommenders}. In \bibinfo{booktitle}{\emph{Proceedings of the 26th ACM
  SIGKDD International Conference on Knowledge Discovery \& Data Mining}}.
  \bibinfo{pages}{483--491}.
\newblock


\bibitem[\protect\citeauthoryear{Oord, Li, and Vinyals}{Oord
  et~al\mbox{.}}{2018}]%
        {oord2018representation}
\bibfield{author}{\bibinfo{person}{Aaron van~den Oord}, \bibinfo{person}{Yazhe
  Li}, {and} \bibinfo{person}{Oriol Vinyals}.} \bibinfo{year}{2018}\natexlab{}.
\newblock \showarticletitle{Representation learning with contrastive predictive
  coding}.
\newblock \bibinfo{journal}{\emph{arXiv preprint arXiv:1807.03748}}
  (\bibinfo{year}{2018}).
\newblock


\bibitem[\protect\citeauthoryear{Pan, Cai, Chen, Chen, and de~Rijke}{Pan
  et~al\mbox{.}}{2020}]%
        {pan2020star}
\bibfield{author}{\bibinfo{person}{Zhiqiang Pan}, \bibinfo{person}{Fei Cai},
  \bibinfo{person}{Wanyu Chen}, \bibinfo{person}{Honghui Chen}, {and}
  \bibinfo{person}{Maarten de Rijke}.} \bibinfo{year}{2020}\natexlab{}.
\newblock \showarticletitle{Star Graph Neural Networks for Session-based
  Recommendation}. In \bibinfo{booktitle}{\emph{Proceedings of the 29th ACM
  International Conference on Information \& Knowledge Management}}.
  \bibinfo{pages}{1195--1204}.
\newblock


\bibitem[\protect\citeauthoryear{Qiao, Shen, Zhang, Wang, and Yuille}{Qiao
  et~al\mbox{.}}{2018}]%
        {qiao2018deep}
\bibfield{author}{\bibinfo{person}{Siyuan Qiao}, \bibinfo{person}{Wei Shen},
  \bibinfo{person}{Zhishuai Zhang}, \bibinfo{person}{Bo Wang}, {and}
  \bibinfo{person}{Alan Yuille}.} \bibinfo{year}{2018}\natexlab{}.
\newblock \showarticletitle{Deep co-training for semi-supervised image
  recognition}. In \bibinfo{booktitle}{\emph{Proceedings of the european
  conference on computer vision (eccv)}}. \bibinfo{pages}{135--152}.
\newblock


\bibitem[\protect\citeauthoryear{Qiu, Chen, Dong, Zhang, Yang, Ding, Wang, and
  Tang}{Qiu et~al\mbox{.}}{2020}]%
        {qiu2020gcc}
\bibfield{author}{\bibinfo{person}{Jiezhong Qiu}, \bibinfo{person}{Qibin Chen},
  \bibinfo{person}{Yuxiao Dong}, \bibinfo{person}{Jing Zhang},
  \bibinfo{person}{Hongxia Yang}, \bibinfo{person}{Ming Ding},
  \bibinfo{person}{Kuansan Wang}, {and} \bibinfo{person}{Jie Tang}.}
  \bibinfo{year}{2020}\natexlab{}.
\newblock \showarticletitle{GCC: Graph Contrastive Coding for Graph Neural
  Network Pre-Training}. In \bibinfo{booktitle}{\emph{Proceedings of the 26th
  ACM SIGKDD International Conference on Knowledge Discovery \& Data Mining}}.
  \bibinfo{pages}{1150--1160}.
\newblock


\bibitem[\protect\citeauthoryear{Qiu, Li, Huang, and Yin}{Qiu
  et~al\mbox{.}}{2019}]%
        {qiu2019rethinking}
\bibfield{author}{\bibinfo{person}{Ruihong Qiu}, \bibinfo{person}{Jingjing Li},
  \bibinfo{person}{Zi Huang}, {and} \bibinfo{person}{Hongzhi Yin}.}
  \bibinfo{year}{2019}\natexlab{}.
\newblock \showarticletitle{Rethinking the Item Order in Session-based
  Recommendation with Graph Neural Networks}. In
  \bibinfo{booktitle}{\emph{Proceedings of the 28th ACM International
  Conference on Information and Knowledge Management}}.
  \bibinfo{pages}{579--588}.
\newblock


\bibitem[\protect\citeauthoryear{Rendle, Freudenthaler, and
  Schmidt-Thieme}{Rendle et~al\mbox{.}}{2010}]%
        {rendle2010factorizing}
\bibfield{author}{\bibinfo{person}{Steffen Rendle}, \bibinfo{person}{Christoph
  Freudenthaler}, {and} \bibinfo{person}{Lars Schmidt-Thieme}.}
  \bibinfo{year}{2010}\natexlab{}.
\newblock \showarticletitle{Factorizing personalized markov chains for
  next-basket recommendation}. In \bibinfo{booktitle}{\emph{Proceedings of the
  19th international conference on World wide web}}. \bibinfo{pages}{811--820}.
\newblock


\bibitem[\protect\citeauthoryear{Shani, Heckerman, and Brafman}{Shani
  et~al\mbox{.}}{2005}]%
        {shani2005mdp}
\bibfield{author}{\bibinfo{person}{Guy Shani}, \bibinfo{person}{David
  Heckerman}, {and} \bibinfo{person}{Ronen~I Brafman}.}
  \bibinfo{year}{2005}\natexlab{}.
\newblock \showarticletitle{An MDP-based recommender system}.
\newblock \bibinfo{journal}{\emph{Journal of Machine Learning Research}}
  \bibinfo{volume}{6}, \bibinfo{number}{Sep} (\bibinfo{year}{2005}),
  \bibinfo{pages}{1265--1295}.
\newblock


\bibitem[\protect\citeauthoryear{Sun, Liu, Wu, Pei, Lin, Ou, and Jiang}{Sun
  et~al\mbox{.}}{2019b}]%
        {sun2019bert4rec}
\bibfield{author}{\bibinfo{person}{Fei Sun}, \bibinfo{person}{Jun Liu},
  \bibinfo{person}{Jian Wu}, \bibinfo{person}{Changhua Pei},
  \bibinfo{person}{Xiao Lin}, \bibinfo{person}{Wenwu Ou}, {and}
  \bibinfo{person}{Peng Jiang}.} \bibinfo{year}{2019}\natexlab{b}.
\newblock \showarticletitle{BERT4Rec: Sequential recommendation with
  bidirectional encoder representations from transformer}. In
  \bibinfo{booktitle}{\emph{Proceedings of the 28th ACM international
  conference on information and knowledge management}}.
  \bibinfo{pages}{1441--1450}.
\newblock


\bibitem[\protect\citeauthoryear{Sun, Hoffmann, Verma, and Tang}{Sun
  et~al\mbox{.}}{2019a}]%
        {sun2019infograph}
\bibfield{author}{\bibinfo{person}{Fan-Yun Sun}, \bibinfo{person}{Jordan
  Hoffmann}, \bibinfo{person}{Vikas Verma}, {and} \bibinfo{person}{Jian Tang}.}
  \bibinfo{year}{2019}\natexlab{a}.
\newblock \showarticletitle{Infograph: Unsupervised and semi-supervised
  graph-level representation learning via mutual information maximization}.
\newblock \bibinfo{journal}{\emph{arXiv preprint arXiv:1908.01000}}
  (\bibinfo{year}{2019}).
\newblock


\bibitem[\protect\citeauthoryear{Sun, Qian, Chen, Liang, Nguyen, and Yin}{Sun
  et~al\mbox{.}}{2020}]%
        {sun2020go}
\bibfield{author}{\bibinfo{person}{Ke Sun}, \bibinfo{person}{Tieyun Qian},
  \bibinfo{person}{Tong Chen}, \bibinfo{person}{Yile Liang},
  \bibinfo{person}{Quoc Viet~Hung Nguyen}, {and} \bibinfo{person}{Hongzhi
  Yin}.} \bibinfo{year}{2020}\natexlab{}.
\newblock \showarticletitle{Where to go next: Modeling long-and short-term user
  preferences for point-of-interest recommendation}. In
  \bibinfo{booktitle}{\emph{Proceedings of the AAAI Conference on Artificial
  Intelligence}}, Vol.~\bibinfo{volume}{34}. \bibinfo{pages}{214--221}.
\newblock


\bibitem[\protect\citeauthoryear{Tan, Xu, and Liu}{Tan et~al\mbox{.}}{2016}]%
        {tan2016improved}
\bibfield{author}{\bibinfo{person}{Yong~Kiam Tan}, \bibinfo{person}{Xinxing
  Xu}, {and} \bibinfo{person}{Yong Liu}.} \bibinfo{year}{2016}\natexlab{}.
\newblock \showarticletitle{Improved recurrent neural networks for
  session-based recommendations}. In \bibinfo{booktitle}{\emph{Proceedings of
  the 1st Workshop on Deep Learning for Recommender Systems}}.
  \bibinfo{pages}{17--22}.
\newblock


\bibitem[\protect\citeauthoryear{Vaswani, Shazeer, Parmar, Uszkoreit, Jones,
  Gomez, Kaiser, and Polosukhin}{Vaswani et~al\mbox{.}}{2017}]%
        {vaswani2017attention}
\bibfield{author}{\bibinfo{person}{Ashish Vaswani}, \bibinfo{person}{Noam
  Shazeer}, \bibinfo{person}{Niki Parmar}, \bibinfo{person}{Jakob Uszkoreit},
  \bibinfo{person}{Llion Jones}, \bibinfo{person}{Aidan~N Gomez},
  \bibinfo{person}{{\L}ukasz Kaiser}, {and} \bibinfo{person}{Illia
  Polosukhin}.} \bibinfo{year}{2017}\natexlab{}.
\newblock \showarticletitle{Attention is All You Need}. In
  \bibinfo{booktitle}{\emph{Advances in neural information processing
  systems}}. \bibinfo{pages}{5998--6008}.
\newblock


\bibitem[\protect\citeauthoryear{Velickovic, Fedus, Hamilton, Li{\`o}, Bengio,
  and Hjelm}{Velickovic et~al\mbox{.}}{2019}]%
        {velickovic2019deep}
\bibfield{author}{\bibinfo{person}{Petar Velickovic}, \bibinfo{person}{William
  Fedus}, \bibinfo{person}{William~L Hamilton}, \bibinfo{person}{Pietro
  Li{\`o}}, \bibinfo{person}{Yoshua Bengio}, {and} \bibinfo{person}{R~Devon
  Hjelm}.} \bibinfo{year}{2019}\natexlab{}.
\newblock \showarticletitle{Deep Graph Infomax.}. In
  \bibinfo{booktitle}{\emph{ICLR (Poster)}}.
\newblock


\bibitem[\protect\citeauthoryear{Wang, Cao, and Wang}{Wang
  et~al\mbox{.}}{2019}]%
        {wang2019survey}
\bibfield{author}{\bibinfo{person}{Shoujin Wang}, \bibinfo{person}{Longbing
  Cao}, {and} \bibinfo{person}{Yan Wang}.} \bibinfo{year}{2019}\natexlab{}.
\newblock \showarticletitle{A survey on session-based recommender systems}.
\newblock \bibinfo{journal}{\emph{arXiv preprint arXiv:1902.04864}}
  (\bibinfo{year}{2019}).
\newblock


\bibitem[\protect\citeauthoryear{Wang, Zhang, Liu, Liu, Zhang, Lin, and
  Zha}{Wang et~al\mbox{.}}{2020b}]%
        {wang2020beyond}
\bibfield{author}{\bibinfo{person}{Wen Wang}, \bibinfo{person}{Wei Zhang},
  \bibinfo{person}{Shukai Liu}, \bibinfo{person}{Qi Liu}, \bibinfo{person}{Bo
  Zhang}, \bibinfo{person}{Leyu Lin}, {and} \bibinfo{person}{Hongyuan Zha}.}
  \bibinfo{year}{2020}\natexlab{b}.
\newblock \showarticletitle{Beyond clicks: Modeling multi-relational item graph
  for session-based target behavior prediction}. In
  \bibinfo{booktitle}{\emph{Proceedings of The Web Conference 2020}}.
  \bibinfo{pages}{3056--3062}.
\newblock


\bibitem[\protect\citeauthoryear{Wang, Wei, Cong, Li, Mao, and Qiu}{Wang
  et~al\mbox{.}}{2020a}]%
        {wang2020global}
\bibfield{author}{\bibinfo{person}{Ziyang Wang}, \bibinfo{person}{Wei Wei},
  \bibinfo{person}{Gao Cong}, \bibinfo{person}{Xiao-Li Li},
  \bibinfo{person}{Xian-Ling Mao}, {and} \bibinfo{person}{Minghui Qiu}.}
  \bibinfo{year}{2020}\natexlab{a}.
\newblock \showarticletitle{Global context enhanced graph neural networks for
  session-based recommendation}. In \bibinfo{booktitle}{\emph{Proceedings of
  the 43rd International ACM SIGIR Conference on Research and Development in
  Information Retrieval}}. \bibinfo{pages}{169--178}.
\newblock


\bibitem[\protect\citeauthoryear{Wu, Zhang, Souza~Jr, Fifty, Yu, and
  Weinberger}{Wu et~al\mbox{.}}{2019b}]%
        {wu2019simplifying}
\bibfield{author}{\bibinfo{person}{Felix Wu}, \bibinfo{person}{Tianyi Zhang},
  \bibinfo{person}{Amauri Holanda~de Souza~Jr}, \bibinfo{person}{Christopher
  Fifty}, \bibinfo{person}{Tao Yu}, {and} \bibinfo{person}{Kilian~Q
  Weinberger}.} \bibinfo{year}{2019}\natexlab{b}.
\newblock \showarticletitle{Simplifying graph convolutional networks}.
\newblock \bibinfo{journal}{\emph{arXiv preprint arXiv:1902.07153}}
  (\bibinfo{year}{2019}).
\newblock


\bibitem[\protect\citeauthoryear{Wu, Lin, Gao, Tan, Li, et~al\mbox{.}}{Wu
  et~al\mbox{.}}{2021}]%
        {wu2021ssg}
\bibfield{author}{\bibinfo{person}{Lirong Wu}, \bibinfo{person}{Haitao Lin},
  \bibinfo{person}{Zhangyang Gao}, \bibinfo{person}{Cheng Tan},
  \bibinfo{person}{Stan Li}, {et~al\mbox{.}}} \bibinfo{year}{2021}\natexlab{}.
\newblock \showarticletitle{Self-supervised on Graphs: Contrastive, Generative,
  or Predictive}.
\newblock \bibinfo{journal}{\emph{arXiv preprint arXiv:2105.07342}}
  (\bibinfo{year}{2021}).
\newblock


\bibitem[\protect\citeauthoryear{Wu, Tang, Zhu, Wang, Xie, and Tan}{Wu
  et~al\mbox{.}}{2019a}]%
        {wu2019session}
\bibfield{author}{\bibinfo{person}{Shu Wu}, \bibinfo{person}{Yuyuan Tang},
  \bibinfo{person}{Yanqiao Zhu}, \bibinfo{person}{Liang Wang},
  \bibinfo{person}{Xing Xie}, {and} \bibinfo{person}{Tieniu Tan}.}
  \bibinfo{year}{2019}\natexlab{a}.
\newblock \showarticletitle{Session-based recommendation with graph neural
  networks}. In \bibinfo{booktitle}{\emph{Proceedings of the AAAI Conference on
  Artificial Intelligence}}, Vol.~\bibinfo{volume}{33}.
  \bibinfo{pages}{346--353}.
\newblock


\bibitem[\protect\citeauthoryear{Wu, Pan, Chen, Long, Zhang, and Philip}{Wu
  et~al\mbox{.}}{2020}]%
        {wu2020comprehensive}
\bibfield{author}{\bibinfo{person}{Zonghan Wu}, \bibinfo{person}{Shirui Pan},
  \bibinfo{person}{Fengwen Chen}, \bibinfo{person}{Guodong Long},
  \bibinfo{person}{Chengqi Zhang}, {and} \bibinfo{person}{S~Yu Philip}.}
  \bibinfo{year}{2020}\natexlab{}.
\newblock \showarticletitle{A comprehensive survey on graph neural networks}.
\newblock \bibinfo{journal}{\emph{IEEE Transactions on Neural Networks and
  Learning Systems}} (\bibinfo{year}{2020}).
\newblock


\bibitem[\protect\citeauthoryear{Xia, Yin, Yu, Wang, Cui, and Zhang}{Xia
  et~al\mbox{.}}{2020}]%
        {xia2020self}
\bibfield{author}{\bibinfo{person}{Xin Xia}, \bibinfo{person}{Hongzhi Yin},
  \bibinfo{person}{Junliang Yu}, \bibinfo{person}{Qinyong Wang},
  \bibinfo{person}{Lizhen Cui}, {and} \bibinfo{person}{Xiangliang Zhang}.}
  \bibinfo{year}{2020}\natexlab{}.
\newblock \showarticletitle{Self-Supervised Hypergraph Convolutional Networks
  for Session-based Recommendation}.
\newblock \bibinfo{journal}{\emph{arXiv preprint arXiv:2012.06852}}
  (\bibinfo{year}{2020}).
\newblock


\bibitem[\protect\citeauthoryear{Xie, Sun, Liu, Gao, Ding, and Cui}{Xie
  et~al\mbox{.}}{2020}]%
        {xie2020contrastive}
\bibfield{author}{\bibinfo{person}{Xu Xie}, \bibinfo{person}{Fei Sun},
  \bibinfo{person}{Zhaoyang Liu}, \bibinfo{person}{Jinyang Gao},
  \bibinfo{person}{Bolin Ding}, {and} \bibinfo{person}{Bin Cui}.}
  \bibinfo{year}{2020}\natexlab{}.
\newblock \showarticletitle{Contrastive Pre-training for Sequential
  Recommendation}.
\newblock \bibinfo{journal}{\emph{arXiv preprint arXiv:2010.14395}}
  (\bibinfo{year}{2020}).
\newblock


\bibitem[\protect\citeauthoryear{Xin, Karatzoglou, Arapakis, and Jose}{Xin
  et~al\mbox{.}}{2020}]%
        {xin2020self}
\bibfield{author}{\bibinfo{person}{Xin Xin}, \bibinfo{person}{Alexandros
  Karatzoglou}, \bibinfo{person}{Ioannis Arapakis}, {and}
  \bibinfo{person}{Joemon~M Jose}.} \bibinfo{year}{2020}\natexlab{}.
\newblock \showarticletitle{Self-Supervised Reinforcement Learning
  forRecommender Systems}.
\newblock \bibinfo{journal}{\emph{arXiv preprint arXiv:2006.05779}}
  (\bibinfo{year}{2020}).
\newblock


\bibitem[\protect\citeauthoryear{Xu, Tao, and Xu}{Xu et~al\mbox{.}}{2013}]%
        {xu2013survey}
\bibfield{author}{\bibinfo{person}{Chang Xu}, \bibinfo{person}{Dacheng Tao},
  {and} \bibinfo{person}{Chao Xu}.} \bibinfo{year}{2013}\natexlab{}.
\newblock \showarticletitle{A survey on multi-view learning}.
\newblock \bibinfo{journal}{\emph{arXiv preprint arXiv:1304.5634}}
  (\bibinfo{year}{2013}).
\newblock


\bibitem[\protect\citeauthoryear{Xu, Zhao, Liu, Sheng, Xu, Zhuang, Fang, and
  Zhou}{Xu et~al\mbox{.}}{2019}]%
        {xu2019graph}
\bibfield{author}{\bibinfo{person}{Chengfeng Xu}, \bibinfo{person}{Pengpeng
  Zhao}, \bibinfo{person}{Yanchi Liu}, \bibinfo{person}{Victor~S Sheng},
  \bibinfo{person}{Jiajie Xu}, \bibinfo{person}{Fuzhen Zhuang},
  \bibinfo{person}{Junhua Fang}, {and} \bibinfo{person}{Xiaofang Zhou}.}
  \bibinfo{year}{2019}\natexlab{}.
\newblock \showarticletitle{Graph contextualized self-attention network for
  session-based recommendation}. In \bibinfo{booktitle}{\emph{Proc. 28th Int.
  Joint Conf. Artif. Intell.(IJCAI)}}. \bibinfo{pages}{3940--3946}.
\newblock


\bibitem[\protect\citeauthoryear{Yao, Yi, Zhiyuan~Cheng, Yu, Chen, Menon, Hong,
  Chi, Tjoa, Kang, et~al\mbox{.}}{Yao et~al\mbox{.}}{2020}]%
        {yao2020self}
\bibfield{author}{\bibinfo{person}{Tiansheng Yao}, \bibinfo{person}{Xinyang
  Yi}, \bibinfo{person}{Derek Zhiyuan~Cheng}, \bibinfo{person}{Felix Yu},
  \bibinfo{person}{Ting Chen}, \bibinfo{person}{Aditya Menon},
  \bibinfo{person}{Lichan Hong}, \bibinfo{person}{Ed~H Chi},
  \bibinfo{person}{Steve Tjoa}, \bibinfo{person}{Jieqi Kang}, {et~al\mbox{.}}}
  \bibinfo{year}{2020}\natexlab{}.
\newblock \showarticletitle{Self-supervised learning for deep models in
  recommendations}.
\newblock \bibinfo{journal}{\emph{arXiv e-prints}} (\bibinfo{year}{2020}),
  \bibinfo{pages}{arXiv--2007}.
\newblock


\bibitem[\protect\citeauthoryear{Yin and Cui}{Yin and Cui}{2016}]%
        {yin2016spatio}
\bibfield{author}{\bibinfo{person}{Hongzhi Yin} {and} \bibinfo{person}{Bin
  Cui}.} \bibinfo{year}{2016}\natexlab{}.
\newblock \bibinfo{booktitle}{\emph{Spatio-temporal recommendation in social
  media}}.
\newblock \bibinfo{publisher}{Springer}.
\newblock


\bibitem[\protect\citeauthoryear{Yu, Gao, Li, Yin, and Liu}{Yu
  et~al\mbox{.}}{2018}]%
        {yu2018adaptive}
\bibfield{author}{\bibinfo{person}{Junliang Yu}, \bibinfo{person}{Min Gao},
  \bibinfo{person}{Jundong Li}, \bibinfo{person}{Hongzhi Yin}, {and}
  \bibinfo{person}{Huan Liu}.} \bibinfo{year}{2018}\natexlab{}.
\newblock \showarticletitle{Adaptive implicit friends identification over
  heterogeneous network for social recommendation}. In
  \bibinfo{booktitle}{\emph{Proceedings of the 27th ACM international
  conference on information and knowledge management}}.
  \bibinfo{pages}{357--366}.
\newblock


\bibitem[\protect\citeauthoryear{Yu, Yin, Gao, Xia, Zhang, and Hung}{Yu
  et~al\mbox{.}}{2021a}]%
        {yu2021socially}
\bibfield{author}{\bibinfo{person}{Junliang Yu}, \bibinfo{person}{Hongzhi Yin},
  \bibinfo{person}{Min Gao}, \bibinfo{person}{Xin Xia},
  \bibinfo{person}{Xiangliang Zhang}, {and} \bibinfo{person}{Nguyen Quoc~Viet
  Hung}.} \bibinfo{year}{2021}\natexlab{a}.
\newblock \showarticletitle{Socially-Aware Self-Supervised Tri-Training for
  Recommendation}.
\newblock \bibinfo{journal}{\emph{arXiv preprint arXiv:2106.03569}}
  (\bibinfo{year}{2021}).
\newblock


\bibitem[\protect\citeauthoryear{Yu, Yin, Li, Gao, Huang, and Cui}{Yu
  et~al\mbox{.}}{2020}]%
        {yu2020enhance}
\bibfield{author}{\bibinfo{person}{Junliang Yu}, \bibinfo{person}{Hongzhi Yin},
  \bibinfo{person}{Jundong Li}, \bibinfo{person}{Min Gao}, \bibinfo{person}{Zi
  Huang}, {and} \bibinfo{person}{Lizhen Cui}.} \bibinfo{year}{2020}\natexlab{}.
\newblock \showarticletitle{Enhance Social Recommendation with Adversarial
  Graph Convolutional Networks}.
\newblock \bibinfo{journal}{\emph{arXiv preprint arXiv:2004.02340}}
  (\bibinfo{year}{2020}).
\newblock


\bibitem[\protect\citeauthoryear{Yu, Yin, Li, Wang, Hung, and Zhang}{Yu
  et~al\mbox{.}}{2021b}]%
        {yu2021self}
\bibfield{author}{\bibinfo{person}{Junliang Yu}, \bibinfo{person}{Hongzhi Yin},
  \bibinfo{person}{Jundong Li}, \bibinfo{person}{Qinyong Wang},
  \bibinfo{person}{Nguyen Quoc~Viet Hung}, {and} \bibinfo{person}{Xiangliang
  Zhang}.} \bibinfo{year}{2021}\natexlab{b}.
\newblock \showarticletitle{Self-Supervised Multi-Channel Hypergraph
  Convolutional Network for Social Recommendation}. In
  \bibinfo{booktitle}{\emph{Proceedings of the Web Conference 2021}}.
  \bibinfo{pages}{413--424}.
\newblock


\bibitem[\protect\citeauthoryear{Zhang, Tang, Zhang, and Xue}{Zhang
  et~al\mbox{.}}{2014b}]%
        {zhang2014addressing}
\bibfield{author}{\bibinfo{person}{Mi Zhang}, \bibinfo{person}{Jie Tang},
  \bibinfo{person}{Xuchen Zhang}, {and} \bibinfo{person}{Xiangyang Xue}.}
  \bibinfo{year}{2014}\natexlab{b}.
\newblock \showarticletitle{Addressing cold start in recommender systems: A
  semi-supervised co-training algorithm}. In
  \bibinfo{booktitle}{\emph{Proceedings of the 37th international ACM SIGIR
  conference on Research \& development in information retrieval}}.
  \bibinfo{pages}{73--82}.
\newblock


\bibitem[\protect\citeauthoryear{Zhang, Dai, Xu, Feng, Wang, Bian, Wang, and
  Liu}{Zhang et~al\mbox{.}}{2014a}]%
        {zhang2014sequential}
\bibfield{author}{\bibinfo{person}{Yuyu Zhang}, \bibinfo{person}{Hanjun Dai},
  \bibinfo{person}{Chang Xu}, \bibinfo{person}{Jun Feng},
  \bibinfo{person}{Taifeng Wang}, \bibinfo{person}{Jiang Bian},
  \bibinfo{person}{Bin Wang}, {and} \bibinfo{person}{Tie-Yan Liu}.}
  \bibinfo{year}{2014}\natexlab{a}.
\newblock \showarticletitle{Sequential click prediction for sponsored search
  with recurrent neural networks}. In \bibinfo{booktitle}{\emph{Twenty-Eighth
  AAAI Conference on Artificial Intelligence}}.
\newblock


\bibitem[\protect\citeauthoryear{Zhang, Yin, Huang, Du, Yang, and Lian}{Zhang
  et~al\mbox{.}}{2018}]%
        {zhang2018discrete}
\bibfield{author}{\bibinfo{person}{Yan Zhang}, \bibinfo{person}{Hongzhi Yin},
  \bibinfo{person}{Zi Huang}, \bibinfo{person}{Xingzhong Du},
  \bibinfo{person}{Guowu Yang}, {and} \bibinfo{person}{Defu Lian}.}
  \bibinfo{year}{2018}\natexlab{}.
\newblock \showarticletitle{Discrete deep learning for fast content-aware
  recommendation}. In \bibinfo{booktitle}{\emph{Proceedings of the eleventh ACM
  international conference on web search and data mining}}.
  \bibinfo{pages}{717--726}.
\newblock


\bibitem[\protect\citeauthoryear{Zhou, Wang, Zhao, Zhu, Wang, Zhang, Wang, and
  Wen}{Zhou et~al\mbox{.}}{2020}]%
        {zhou2020s}
\bibfield{author}{\bibinfo{person}{Kun Zhou}, \bibinfo{person}{Hui Wang},
  \bibinfo{person}{Wayne~Xin Zhao}, \bibinfo{person}{Yutao Zhu},
  \bibinfo{person}{Sirui Wang}, \bibinfo{person}{Fuzheng Zhang},
  \bibinfo{person}{Zhongyuan Wang}, {and} \bibinfo{person}{Ji-Rong Wen}.}
  \bibinfo{year}{2020}\natexlab{}.
\newblock \showarticletitle{S\^{} 3-Rec: Self-Supervised Learning for
  Sequential Recommendation with Mutual Information Maximization}.
\newblock \bibinfo{journal}{\emph{arXiv preprint arXiv:2008.07873}}
  (\bibinfo{year}{2020}).
\newblock


\bibitem[\protect\citeauthoryear{Zimdars, Chickering, and Meek}{Zimdars
  et~al\mbox{.}}{2013}]%
        {zimdars2013using}
\bibfield{author}{\bibinfo{person}{Andrew Zimdars},
  \bibinfo{person}{David~Maxwell Chickering}, {and}
  \bibinfo{person}{Christopher Meek}.} \bibinfo{year}{2013}\natexlab{}.
\newblock \showarticletitle{Using temporal data for making recommendations}.
\newblock \bibinfo{journal}{\emph{arXiv preprint arXiv:1301.2320}}
  (\bibinfo{year}{2013}).
\newblock


\end{thebibliography}
\end{document}